%% file: HIG-19-012_temp.tex
\pdfoutput=1

\documentclass[11pt,twoside,a4paper,cmspaper,final,collab]{cms-tdr}
\def\svnVersion{d7e8e58-D}\def\svnDate{2020/11/12}\def\cmsCernNoTag{CERN-EP-2020-120}\def\cmsCernDate{\today}\def\cmsMessage{"Published in the Journal of High Energy Physics as \href{http://dx.doi.org/10.1007/JHEP11(2020)039}{\doi{10.1007/JHEP11(2020)039}.}"}

\begin{document}\cmsNoteHeader{HIG-19-012}

\hyphenation{had-ron-i-za-tion}
\hyphenation{cal-or-i-me-ter}
\hyphenation{de-vices}
\newcommand{\Irelm}{\ensuremath{I_{\text{rel}}^{\Pgm}}}
\newcommand{\Itrk}{\ensuremath{I^{\text{trk}}}}
\newcommand{\mumurho}{\Pgm\Pgm\Pgp\Pgp}
\newcommand{\mumuphi}{\Pgm\Pgm\PK\PK}
\newcommand{\eerho}{\Pe\Pe\Pgp\Pgp}
\newcommand{\eephi}{\Pe\Pe\PK\PK}
\newcommand{\ee}{\Pe\Pe}
\newcommand{\mumu}{\Pgm\Pgm}
\newcommand{\llrho}{\ell\ell\Pgp\Pgp}
\newcommand{\mllhad}{m_{\ell\ell\mathrm{h}\mathrm{h}}}
\newcommand{\llphi}{\ell\ell\PK\PK}
\newcommand{\Hzrho}{\PH\to\PZ\PGr}
\newcommand{\Hzphi}{\PH\to\PZ\PGf}
\newcommand{\zphi}{\PZ\PGf}
\newcommand{\zrho}{\PZ\PGr}
\newcommand{\ztoll}{\PZ\to\ell\ell}
\newcommand{\ggH}{\Pg\Pg\PH}
\newcommand{\WH}{\PW\PH}
\newcommand{\ZH}{\PZ\PH}
\newcommand{\zz}{\PZ\PZ^{*}}
\newcommand{\Hzz}{\PH\to\zz}
\newcommand{\ppc}{\Pp\Pp}
\newcommand{\mpipi}{m_{\Pgp\Pgp}}
\newcommand{\pippim}{\Pgpp\Pgpm}
\newcommand{\kpkm}{\PKp\PKm}
\newcommand{\mkk}{m_{\PK\PK}}
\newcommand{\ea}{\epsilon\mathcal{A}}
\newcommand{\dbias}{\Delta_{\text{bias}}}
\newcommand{\mumupm}{\Pgmp\Pgmm}
\newcommand{\pgjpsi}{\PGg\PJGy}
\newcommand{\pgy}{\ensuremath{\PGg\PGy}(2S)\xspace}
\newcommand{\pgu}{\ensuremath{\PGg\PGU}(nS)\xspace}
\newcommand{\pgr}{\ensuremath{\PGg\PGr}\xspace}
\newcommand{\pgf}{\ensuremath{\PGg\PGf}\xspace}
\newlength\cmsTabSkip\setlength{\cmsTabSkip}{1ex}
\providecommand{\cmsTable}[1]{\resizebox{\textwidth}{!}{#1}}
\renewcommand{\labelenumi}{\theenumi}

\cmsNoteHeader{HIG-19-012}
\title{Search for decays of the 125\GeV Higgs boson into a \PZ boson and a \PGr or \PGf meson}

\author*[inst1]{Adinda De Wit}

\date{\today}

\abstract{
Decays of the 125\GeV Higgs boson into a \PZ boson and a \PGrzP{770} or \Pgf meson are searched for using proton-proton collision data collected by the CMS experiment at the LHC at $\sqrt{s}=13\TeV$.
The analysed data set corresponds to an integrated luminosity of 137\fbinv. Events are selected in which the \PZ boson decays into a pair of electrons or a pair of muons, and the \PGr and \PGf mesons decay into pairs of pions and kaons, respectively. No significant excess above the background model is observed. As different polarization states are possible for the decay products of the \PZ boson and \PGr or \PGf mesons, affecting the signal acceptance, scenarios in which the decays are longitudinally or transversely polarized are considered. Upper limits at the 95\% confidence level on the Higgs boson branching fractions into $\zrho$ and $\zphi$ are determined to be 1.04--1.31\% and 0.31--0.40\%, respectively, where the ranges reflect the considered polarization scenarios; these values are 740--940 and 730--950 times larger than the respective standard model expectations. These results constitute the first experimental limits on the two decay channels.}

\hypersetup{%
pdfauthor={CMS Collaboration},%
pdftitle={Search for decays of the 125 GeV Higgs boson into a Z boson and a rho or phi meson},%
pdfsubject={CMS},%
pdfkeywords={CMS, physics, Higgs, rare decays}}

\maketitle

\section{Introduction}
\label{sec:introduction}

In 2012 a boson with a mass near 125\GeV was discovered by the ATLAS and CMS Collaborations at the CERN LHC~\cite{Aad:2012tfa,Chatrchyan:2012xdj,Chatrchyan:2013lba}. 
Soon after it was established that the properties of this particle are, within uncertainties, in agreement with
those of the Higgs boson (\PH) in the standard model (SM)~\cite{Englert:1964et,Higgs:1964ia,Higgs:1964pj,Guralnik:1964eu,Higgs:1966ev,Kibble:1967sv}. 
Decays of the Higgs boson into $\Pgg\Pgg$, $\zz$, $\PWpm\PWmp^{*}$, $\PGtp\PGtm$, and \bbbar, as well as Higgs boson production via gluon-gluon fusion ($\ggH$), via vector boson fusion (VBF), in association with a vector boson, and in association with a top quark-antiquark pair, have all been observed~\cite{Aad:2014eha,Khachatryan:2014ira,Aad:2014eva,Chatrchyan:2013mxa,ATLAS:2014aga,Aad:2015ona,Chatrchyan:2013iaa,Khachatryan:2016vau,Aaboud:2018pen,Sirunyan:2017khh,Aaboud:2018urx,Sirunyan:2018hoz,Aaboud:2018zhk,Sirunyan:2018kst}.
While many of the couplings between the Higgs boson and other particles have already been measured, the required
sensitivity for measuring Yukawa couplings to second- and first-generation fermions has not yet been reached. 
Yukawa couplings to second-generation fermions are accessible via searches for the decay of the Higgs boson into $\mumupm$ or \ccbar, both of which have been performed at the LHC~\cite{Aad:2020xfq,Sirunyan:2018hbu,Aaboud:2018fhh,Sirunyan:2019qia}. The upper limit at the 95\% confidence level (\CL) for the decay into $\mumupm$ (\ccbar) is approximately 2 (70) times the SM expectation. In addition, Yukawa couplings to 
lighter fermions are also accessible via rare exclusive decays of the Higgs boson. One class of such processes is the decay of the Higgs boson into a photon and a vector meson~\cite{Bodwin:2013gca,Kagan:2014ila,Alte:2016yuw}. Thus far, the 
$\pgjpsi$, \pgy, \pgu, $\pgr$, and $\pgf$ decays 
have been searched for~\cite{Sirunyan:2018fmm,Aaboud:2018txb,Aaboud:2017xnb}. The 95\% \CL upper limits on the branching fractions of the Higgs boson into $\pgjpsi$, $\pgr$, and $\pgf$ are 2 orders of magnitude larger than their expected values in the SM. For the \pgy and \pgu decays, the corresponding upper limits are, respectively, 3 and 5 orders of magnitude larger than the SM expectation.

\begin{figure}[h!]
  \centering
    \includegraphics[width=0.3\textwidth]{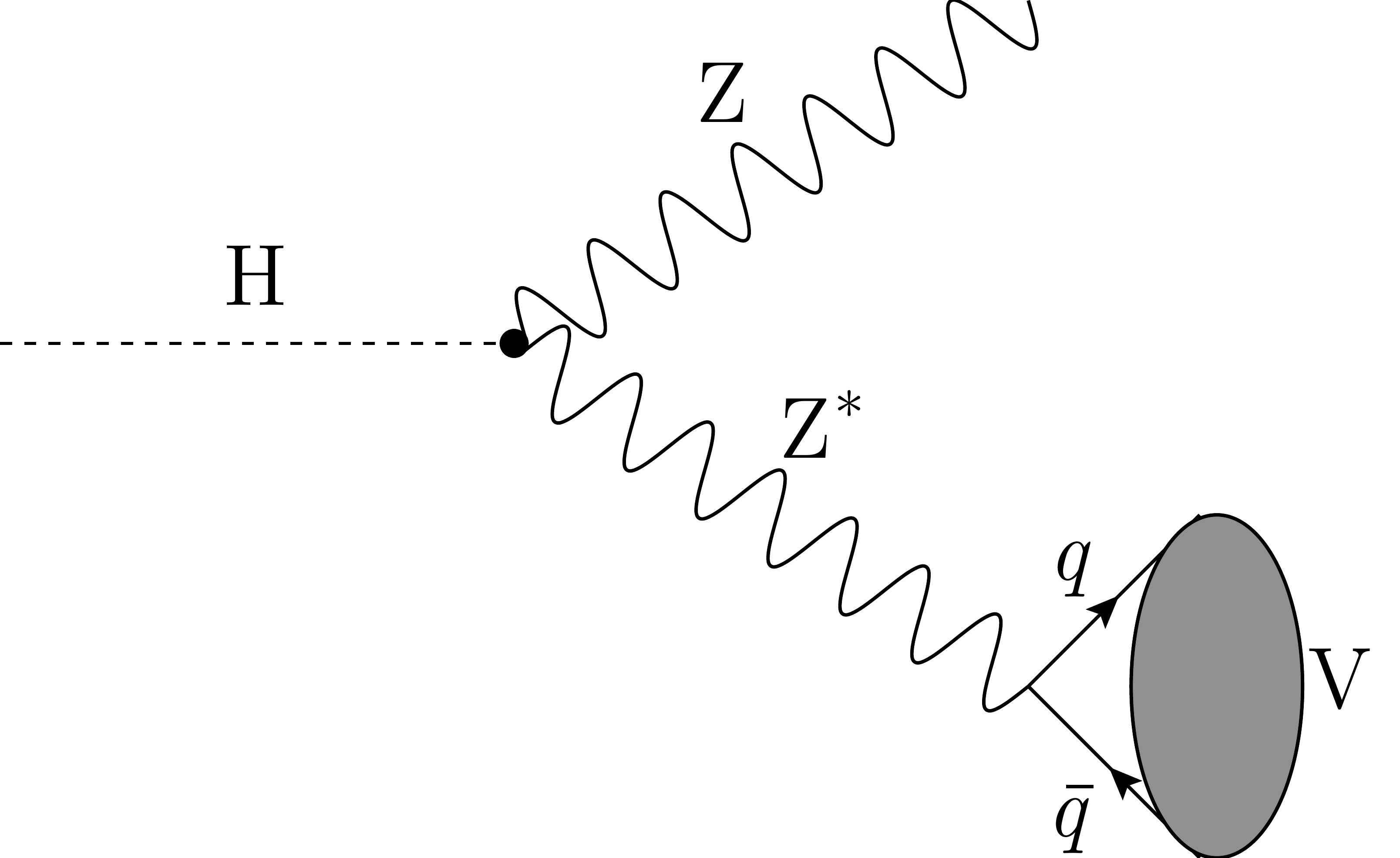}
    \includegraphics[width=0.3\textwidth]{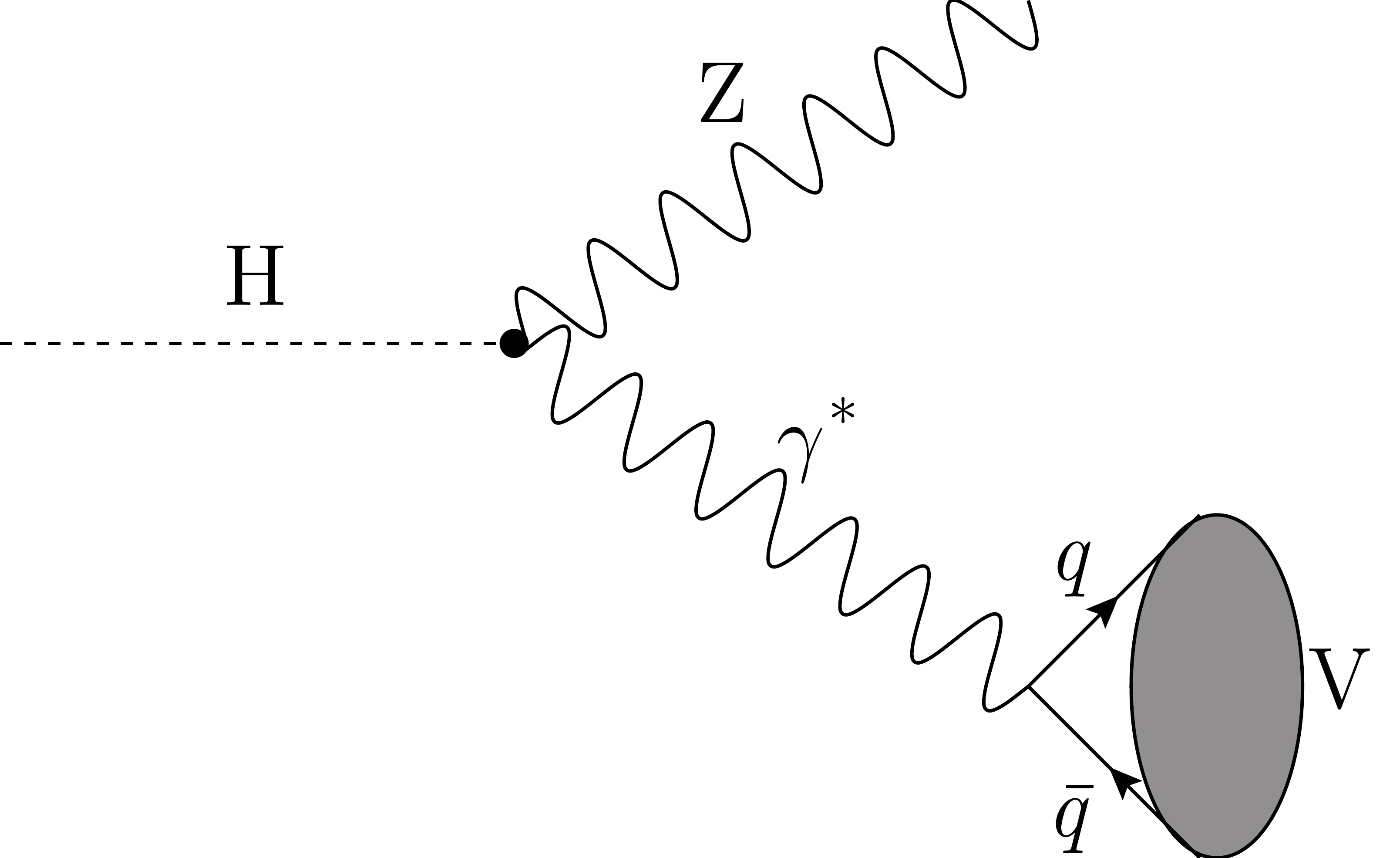}
    \includegraphics[width=0.3\textwidth]{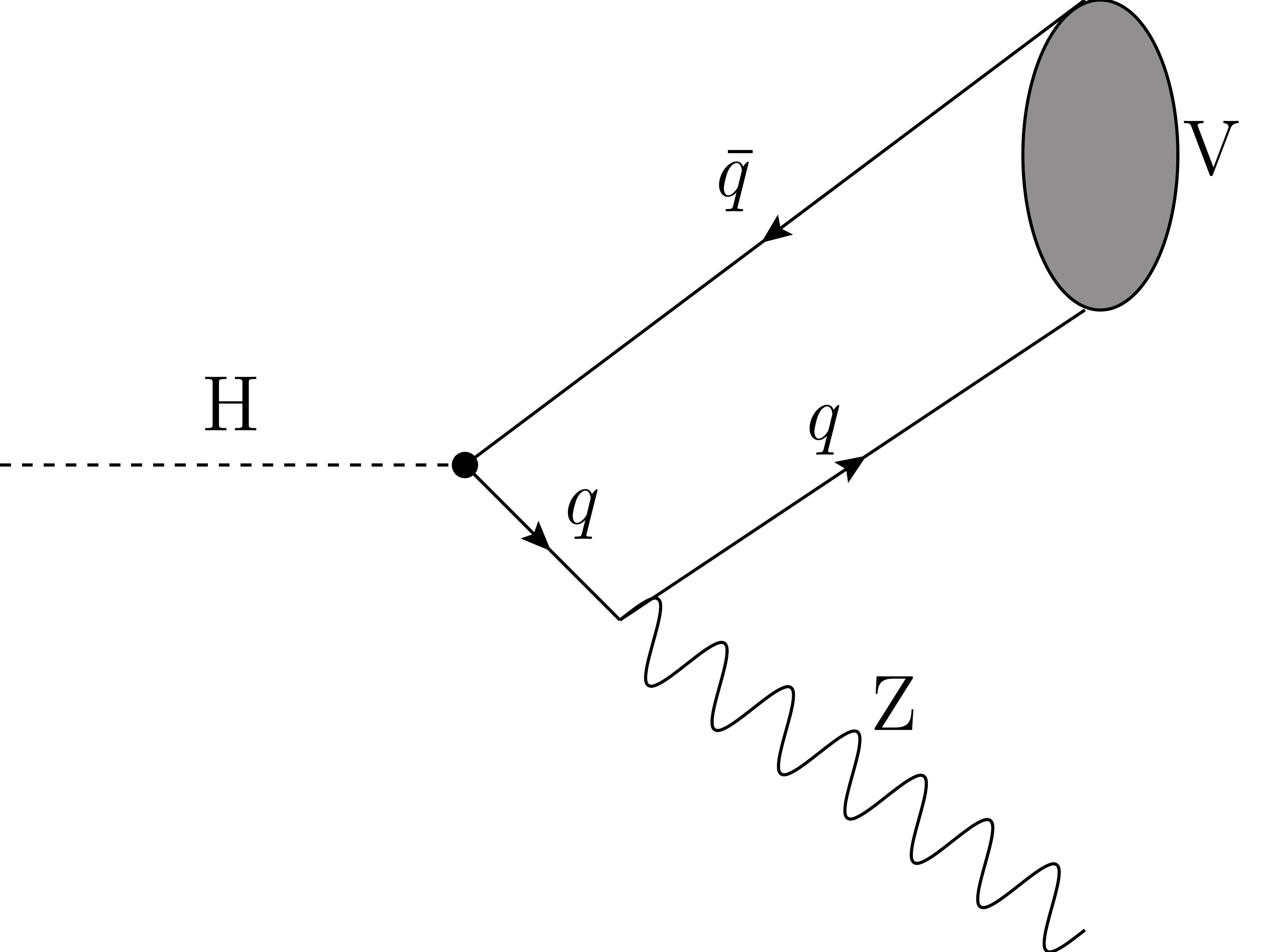}
  \caption{Feynman diagrams of processes that contribute to the decay of a Higgs boson into a heavy vector boson and a vector meson. The grey oval shape represents the meson. The two indirect processes (left and middle), where the meson originates from an off-shell \PZ boson or $\PGg^{*}$, 
contribute the most to the total branching fraction in the SM. }
  \label{fig:diagrams}
\end{figure}

A related class of rare decays is that of the Higgs boson into a heavy vector boson and a vector meson (V)~\cite{deFlorian:2016spz,Isidori:2013cla}. Up to now only the decays of the Higgs boson into $\PZ\PJGy$ and $\PZ\PGhc$ have been studied experimentally~\cite{Aad:2020hzm}. As indicated in Fig.~\ref{fig:diagrams}, several processes contribute to the decay of the Higgs boson into a vector boson and a meson. The formation of a vector boson and a meson via $\Hzz$ or $\PH\to\PZ\PGg^{*}$ decays (Fig.~\ref{fig:diagrams}, left and middle) are indirect contributions to this process. We refer to the decay of the Higgs boson into light quarks that radiate a vector boson and form a bound meson state (Fig.~\ref{fig:diagrams}, right) as the direct process. In the SM the indirect processes contribute the most to the decay of the Higgs boson into a heavy vector boson and a vector meson. 

The direct process is negligible in the SM as it is suppressed by a factor of up to $m_{\PQq}^2/m_{\PH}^2$ relative to the indirect contributions~\cite{Alte:2016yuw}. In that expression $m_{\PQq}$ and $m_{\PH}$ denote the masses of the quark and of the Higgs boson, respectively. However, in scenarios beyond the SM where the Yukawa couplings to light fermions are enhanced, this direct process could contribute significantly to
the Higgs boson branching fraction into a vector boson and a meson~\cite{deFlorian:2016spz}. An 
example of a model beyond the SM with enhanced Yukawa couplings to light fermions is a version of the Giudice--Lebedev model of quark masses~\cite{Giudice:2008uua} that is modified
to have two Higgs doublets. In this scenario Yukawa couplings to light quarks could be enhanced by up to a factor of 7~\cite{Bishara:2015cha}. Enhancements of the Yukawa couplings by factors up to 30, 500, and 2000 for, respectively, strange, down, and up quarks are possible in a two Higgs doublet model with spontaneous flavour violation~\cite{Egana-Ugrinovic:2019dqu}. Other scenarios in which light-quark Yukawa couplings
can be larger than predicted in the SM include a single Higgs doublet model with Froggatt--Nielsen mechanism~\cite{Froggatt:1978nt} and Randall--Sundrum models of warped extra dimensions~\cite{Randall:1999ee,Huber:2000ie}. In addition, studies of the indirect processes are also of interest as these  probe a different phase space from conventional $\PH\to\PW\PW^{*}$ and $\Hzz$ measurements, and therefore provide complementary information.

This paper describes a search for decays of the 125\GeV Higgs boson into a
\PZ boson and a $\Pgr^0$ meson ($\Hzrho$) or into a \PZ boson and a \Pgf meson ($\Hzphi$). 
The branching fractions of these processes in the SM are small: $\mathcal{B}(\Hzrho)=(1.4\pm0.1)\times10^{-5}$ and $\mathcal{B}(\Hzphi)=(4.2\pm0.3)\times10^{-6}$~\cite{deFlorian:2016spz}.
The search uses a sample of proton-proton ($\ppc$) collisions collected by the CMS experiment at $\sqrt{s}=13\TeV$ from 2016 to 2018. 
The data set corresponds to an integrated luminosity of 137\fbinv, or 35.9, 41.5, and 59.7\fbinv collected in 2016, 2017, and 2018, respectively. In this search we select the dimuon and dielectron final states of the \PZ boson. For the \PGr and \PGf mesons, we select
decays containing exactly two charged hadrons, corresponding to the $\pippim$ final state for the \PGr meson and the $\kpkm$ final state for the \PGf meson. In the event 
reconstruction \PGppm and \PKpm are not explicitly distinguished. The main source of background events in this analysis is from Drell--Yan production of a \PZ boson in association with a genuine or misidentified meson candidate.
For brevity we do not distinguish between particles and antiparticles in our notations of decay processes in the remainder of this paper.

\section{The CMS detector}
\label{sec:detector}

The central feature of the CMS apparatus is a superconducting solenoid of 6\unit{m} internal diameter, providing a magnetic field of 3.8\unit{T}. 
Within the solenoid volume are a silicon pixel and strip tracker, a lead tungstate crystal electromagnetic calorimeter (ECAL), and a brass and scintillator hadron calorimeter, each composed of a barrel and two endcap sections. 
Forward calorimeters extend the pseudorapidity ($\eta$) coverage provided by the barrel and endcap detectors. 
Muons are detected in gas-ionization chambers embedded in the steel flux-return yoke outside the solenoid.

The silicon tracker measures charged particles within the pseudorapidity range $\abs{\eta} < 2.5$. It consists of 1856 silicon pixel and 15\,148 silicon strip detector modules. The silicon pixel detector modules
are arranged in four layers. In 2016, data were taken
with a different detector configuration; at that time there were 1440 silicon pixel detector modules arranged in three layers.
 For nonisolated particles with transverse momentum in the range $1 < \pt < 10\GeV$ and $\abs{\eta} < 1.4$, the track resolution is typically 1.5\% in \pt~\cite{TRK-11-001}.

Muons are measured in the pseudorapidity range $\abs{\eta} < 2.4$, with detection planes made using three technologies: drift tubes, cathode strip chambers, and resistive plate chambers. The single-muon trigger efficiency exceeds 90\% over the full $\eta$ range, and the efficiency to reconstruct and identify muons is greater than 96\%. Matching muons to tracks measured in the silicon tracker results in a relative \pt resolution, for muons with \pt up to 100\GeV, of 1\% in the barrel and 3\% in the endcaps~\cite{Sirunyan:2018}.

The electron momentum is estimated by combining the energy measurement in the ECAL with the momentum measurement in the tracker. The momentum resolution for electrons with \mbox{$\pt \approx 45\GeV$} from $\PZ \to \Pe \Pe$ decays ranges from 1.7 to 4.5\%. It is generally better in the barrel region than in the endcaps, and also depends on the bremsstrahlung energy emitted by the electron as it traverses the material in front of the ECAL~\cite{Khachatryan:2015hwa}.

Events of interest are selected using a two-tiered trigger system~\cite{Khachatryan:2016bia}. The first level, composed of custom hardware processors, uses information from the calorimeters and muon detectors to select events at a rate of around 100\unit{kHz} within a time interval of less than 4\mus. The second level, known as the high-level trigger, consists of a farm of processors running a version of the full event reconstruction software optimized for fast processing, and reduces the event rate to around 1\unit{kHz} before data storage.

A more detailed description of the CMS detector, together with a definition of the coordinate system used and the relevant kinematic variables, can be found in Ref.~\cite{Chatrchyan:2008zzk}.

\section{Event reconstruction}
\label{sec:event-reconstruction}
The products of $\ppc$ collisions are reconstructed based on a particle-flow algorithm~\cite{Sirunyan:2017ulk}, which combines information from all subdetectors to reconstruct individual particle 
candidates. These particle candidates are classified as muons, electrons, photons, and charged and neutral hadrons.

The candidate vertex with the largest value of summed physics-object $\pt^2$ is taken to be the primary $\ppc$ interaction vertex (PV). The physics objects are the jets, clustered using the jet finding algorithm~\cite{Cacciari:2008gp,Cacciari:2011ma} with the tracks assigned to candidate vertices as inputs, and the associated missing transverse momentum, taken as the negative vector sum of the \pt of those jets.
Other collision vertices in the event
are considered to have originated from additional inelastic $\ppc$ collisions in each bunch 
crossing, referred to as pileup (PU). The average number of PU interactions during the 2016 data-taking period was 23, rising to 32 during the 2017 and 2018 data-taking periods.
The muons, electrons, and charged hadron tracks used in the search presented in this paper are all required to originate from the PV.

Muons are reconstructed through a simultaneous track fit
to hits in the tracker and in the muon chambers~\cite{Sirunyan:2018}. To suppress
particles misidentified as muons, additional requirements are applied on the track
fit quality and compatibility of individual track segments with the fitted track.
Contamination  
from muons produced within jets is reduced further by requiring the muon to be 
isolated from hadronic activity in the detector. 
A relative isolation variable is defined as
\begin{linenomath}
\begin{equation}
\Irelm=\frac{\sum_{\text{charged}} \pt + \max\Bigl(0,\sum_{\text{neutral}} \pt -0.5 \sum_{\text{charged}} \pt^{\text{PU}}\Bigr)}{\pt^{\Pgm}},
\label{eqn:reliso}
\end{equation}
\end{linenomath}
where $\sum_{\text{charged}} \pt$ refers to the scalar sum of the transverse momenta of all charged particles and $\sum_{\text{neutral}} \pt$ is 
the sum of the \pt of neutral hadrons and photons.
These two sums are calculated within a cone of radius $\Delta R = 0.4$ around the direction of the muon, where $\Delta R = \sqrt{\smash[b]{\left(\Delta\eta\right)^{2}+\left(\Delta\phi\right)^{2}}}$ and 
$\Delta\eta$ and $\Delta \phi$ are differences in pseudorapidity and azimuthal angle, respectively. The \pt of the muon is excluded from these sums. 
To reduce the effects from PU, charged particles are only considered in the isolation sum if they are associated with the PV.
The term $0.5 \sum_{\text{charged}} \pt^{\text{PU}}$ estimates the contributions from neutral particles
in PU by summing the \pt of charged particles that are within the isolation cone but are not associated with the PV.
The factor 0.5 accounts for the ratio of neutral to charged particle production. 
Muons selected in the analysis must satisfy $\Irelm<0.15$. 
After these identification and isolation requirements are imposed, prompt muons are identified with an efficiency of over 90\%. 
A looser selection, 
where the isolation requirement is removed, is also used in the analysis to reject events with additional muons. 

Electrons are reconstructed by combining clusters of energy deposits in the ECAL 
with hits in the tracker~\cite{Khachatryan:2015hwa}. To reduce contamination from particles
incorrectly identified as electrons, reconstructed electrons are required to pass a
multivariate electron identification discriminant. This discriminant, based on the one described in Ref.~\cite{Khachatryan:2015hwa}, 
combines information
about the quality of the tracks, the shower shape, kinematic quantities, and hadronic activity
in the vicinity of the reconstructed electron. Isolation sums similar to those in Eq.~(\ref{eqn:reliso}) are  
also included among the discriminant inputs. Therefore no additional isolation requirements are applied.
Using the requirements placed on the discriminant in this analysis, the
electron identification efficiency is 80\%. The rate at which other particles are 
misidentified as electrons is $\approx$1\%. Looser requirements are used 
to reject events with additional electrons. Using this looser selection
on the multivariate identification discriminant, the electron identification efficiency is 90\% and other particles are misidentified as electrons at a rate of 2--5\%.

The \PGr and \PGf meson decay products are reconstructed using charged particle tracks measured
in the tracker. The tracks are required to originate from 
the PV and to pass ``high purity'' reconstruction requirements.
These requirements are based on the number of tracker layers with hits, 
the track fit quality, and the values of the impact parameters relative to their uncertainties. The algorithm
is described in more detail in Ref.~\cite{TRK-11-001}.
In the event selection, described in Section~\ref{sec:event-selection}, we exploit the known masses of pions and kaons
to calculate and restrict the invariant mass of the \PGr and \PGf candidates. 

\section{Simulated samples}
\label{sec:signal-simulation}
Samples of simulated Higgs boson events, produced via the $\ggH$, VBF, \PW-associated ($\WH$), and
\PZ-associated ($\ZH$) modes, are generated at next-to-leading order (NLO) in quantum chromodynamics (QCD) using \POWHEG 2.0~\cite{Nason:2004rx,Frixione:2007vw,Alioli:2008tz,
Alioli:2010xd,Nason:2009ai,Luisoni:2013kna}. In some of the figures in this paper, and for the evaluation of corrections that account for differences between
data and simulation, samples of simulated Drell--Yan $\ztoll$ events are used. Here, $\ell$ refers to $\Pe\text{ or }\Pgm$.
These samples are generated at leading order using \MGvATNLO 2.2.2 (2.4.2) ~\cite{Alwall:2014hca} for the 2016 (2017 and 2018) data-taking periods. 
All generated samples are interfaced with \PYTHIA 8.212~\cite{Sjostrand:2014zea} to model parton showering and hadronization. In the signal samples the decays $\Hzrho$ or $\Hzphi$ are also modelled using \PYTHIA. These samples are used to build the signal model, which consists of binned templates.
The NNPDF3.0 parton distribution functions (PDFs)~\cite{Ball:2014uwa} are used for the 2016 data-taking period. For the samples of signal events NLO PDFs are used, while
for the Drell--Yan events leading order PDFs are used. For the 2017 and 2018 data-taking periods the NNPDF3.1 PDFs~\cite{Ball:2017nwa} at next-to-next-to-leading order are used for all samples. The description
of the underlying event is provided by the CUETP8M1 tune~\cite{Khachatryan:2015pea} for the 2016 data-taking period and by the CP5 tune~\cite{Sirunyan:2019dfx} for the 
2017 and 2018 data-taking periods.
Additional PU interactions, generated with \PYTHIA, are added to all simulated events
in accordance with the expected PU distribution. All generated events are passed through a \GEANTfour-based~\cite{Agostinelli:2002hh} simulation
of the CMS detector before being reconstructed with the same version of the CMS event
reconstruction software as used for data.

\section{Event selection}
\label{sec:event-selection}
The final states considered in the selection are the $\mumurho$ and $\eerho$ decays of the $\zrho$ system, and the $\mumuphi$ and $\eephi$ decays of the $\zphi$ system.
The selection of the $\mumu$ and $\ee$ pairs, referred to as the dilepton system in what follows, is independent of the meson candidate under consideration.
The trigger selection for the $\mumu$ final states is based on the presence of at least one
isolated muon with $\pt>24\GeV$ in the 2016 and 2018 data-taking periods, and at least one isolated muon with $\pt>27\GeV$ in the 2017 data-taking period. 
For the $\ee$ final states the trigger selection requires the presence of at least one isolated electron with \mbox{$\pt>27\GeV$} in the 2016 data-taking period. In the 2017 (2018) data-taking period this threshold is \mbox{$\pt>35\text{ }(32)\GeV$}.

After imposing the trigger requirements, events in the $\mumu$ channel are selected by requiring the
presence of two oppositely charged muons passing the identification and isolation criteria described in 
Section~\ref{sec:event-reconstruction}. At least one of these muons must pass the trigger selection. Both muons must
have $\pt>20\GeV$ and $\abs{\eta}<2.4$, while the \pt of the muon that satisfies the trigger requirements must be at least 3\GeV above the \pt 
threshold at the trigger level. 
The $\ee$ channel selects events containing two oppositely charged electrons passing the identification criteria described in Section~\ref{sec:event-reconstruction}. At least one of the electrons must pass the trigger selection. Both electrons must have $\pt>20\GeV$ and $\abs{\eta}<2.1$. The \pt 
of the electron satisfying the trigger requirement must be at least 3\GeV above the trigger-level threshold.
The requirement that the \pt of the lepton passing the trigger selection is at least 3\GeV above the threshold in the trigger ensures we
avoid the part of the phase space where the trigger efficiency increases rapidly.
In both the $\mumu$ and $\ee$ channels, events that contain additional leptons with $\pt>5\GeV$ that pass the loose identification
criteria described in Section~\ref{sec:event-reconstruction} are rejected.
The invariant mass of the dilepton system is required to be in the range $60<m_{\ell\ell}<120\GeV$.

\begin{figure}[h!]
  \centering
    \includegraphics[width=0.48\textwidth]{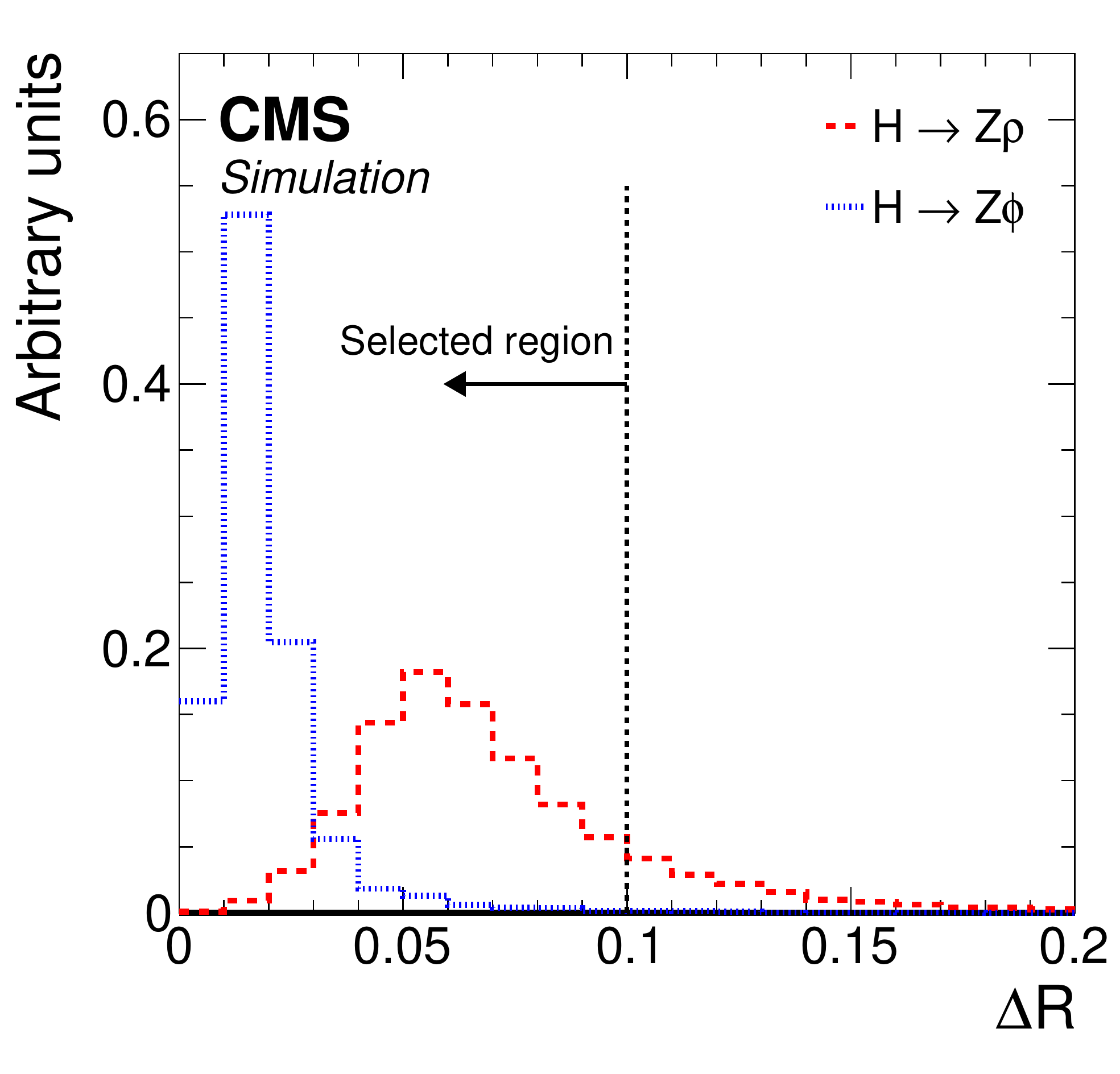}
  \caption{The angular distance $\Delta R$ between the two tracks from the meson decay in $\Hzrho$ events (dashed red) and in $\Hzphi$ events (dotted blue).
  The separation is calculated between reconstructed tracks that are matched to the generator-level pions (kaons) to ensure that the tracks originate from the \PGr (\PGf) decay. Both
contributions are normalized to the same area.
 }
  \label{fig:track_dr}
\end{figure}

\begin{figure}[h!]
  \centering
    \includegraphics[width=0.48\textwidth]{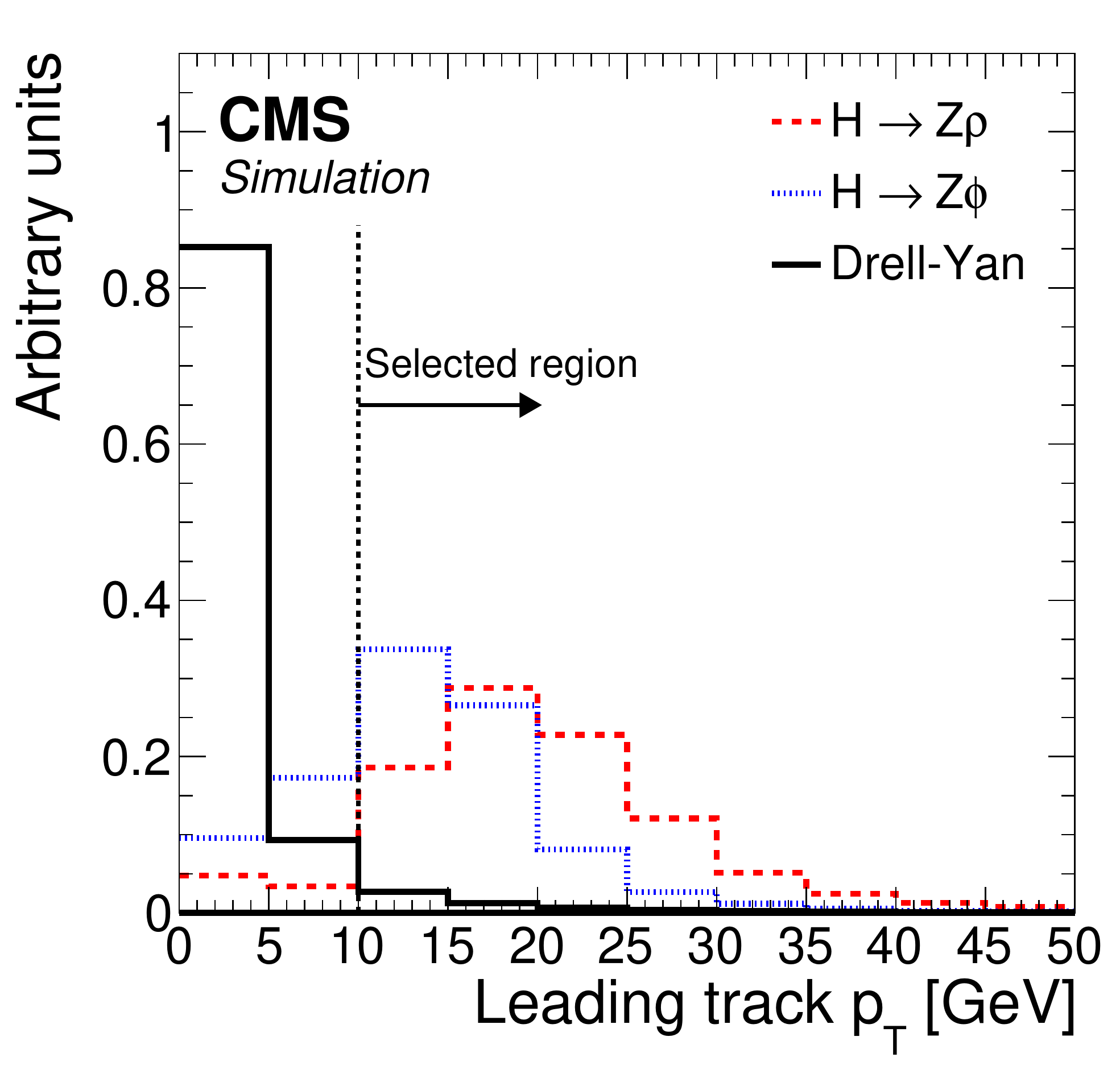}
  \caption{The transverse momentum distribution for the track that has the larger \pt out of the two tracks selected as the \PGr or \PGf candidate. The distribution is shown for events that pass the meson candidate
selection described in the text, but not the requirement that one of the tracks must have $\pt>10\GeV$. This 
distribution is shown for the $\Hzrho$ decay (dashed red), for the $\Hzphi$ decay (dotted blue), and for the background from Drell--Yan events (solid black). All contributions are normalized to the same area.
 }
  \label{fig:track_pt}
\end{figure}

The \PGr (\PGf) candidate is reconstructed from its decay into $\pippim$ ($\kpkm$).
As the \PGr and \PGf mesons are both light compared to the energy released in the decay, the two 
charged particles produced in the decay are emitted with small angular separation $\Delta R$, as illustrated in Fig.~\ref{fig:track_dr}. 
The events shown in this figure are required to pass the selection criteria described so far. 
The small separation between the two tracks is exploited in the selection of the \PGr and \PGf candidates. The meson candidate is selected as a pair of oppositely charged particle tracks, both with $\pt>1\GeV$ and 
separated by $\Delta R < 0.1$. In what follows a pair of oppositely charged particle tracks is also referred to as a ditrack system. The charged particle tracks are required to be separated from each of the \PZ boson decay products by $\Delta R>0.3$.
In addition, at least one of the tracks must have $\pt>10\GeV$. Figure~\ref{fig:track_pt} shows the \pt distribution for the track that has the larger transverse momentum out of the two tracks selected as the meson candidate. This distribution is shown in the $\Hzrho$ and $\Hzphi$ signal events and in the background from Drell--Yan events, illustrating how this requirement helps to reduce the background. If multiple track pairs pass these requirements, we calculate the four-momentum of each ditrack system and select the pair of tracks with the highest \pt. This choice
maximizes the proportion of signal events in which the correct meson candidate is selected.
In all channels, the meson candidate is correctly identified in 98--99\% of the signal events.

\begin{figure}[h!]
\centering
\includegraphics[width=0.48\textwidth]{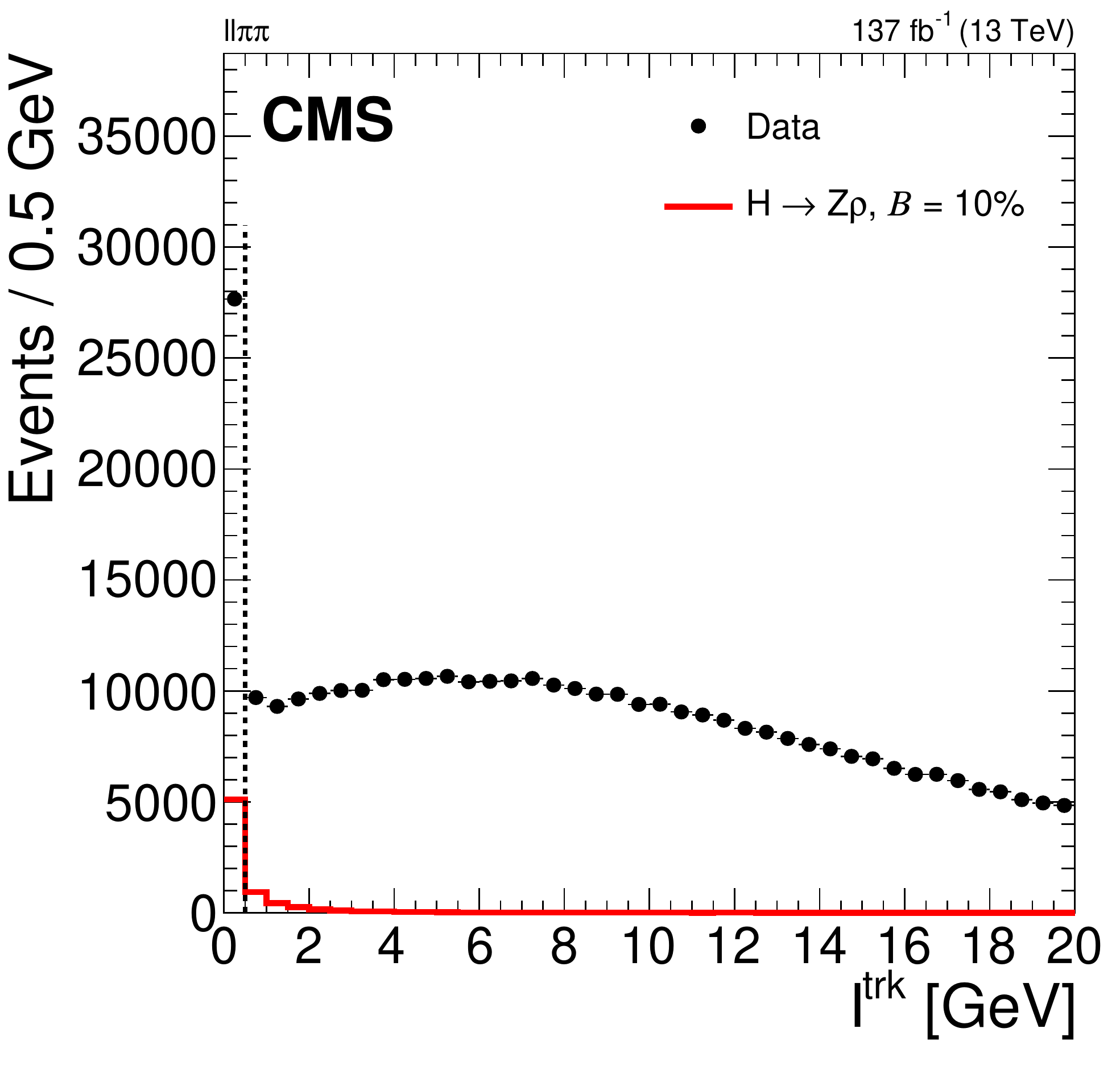}
\includegraphics[width=0.48\textwidth]{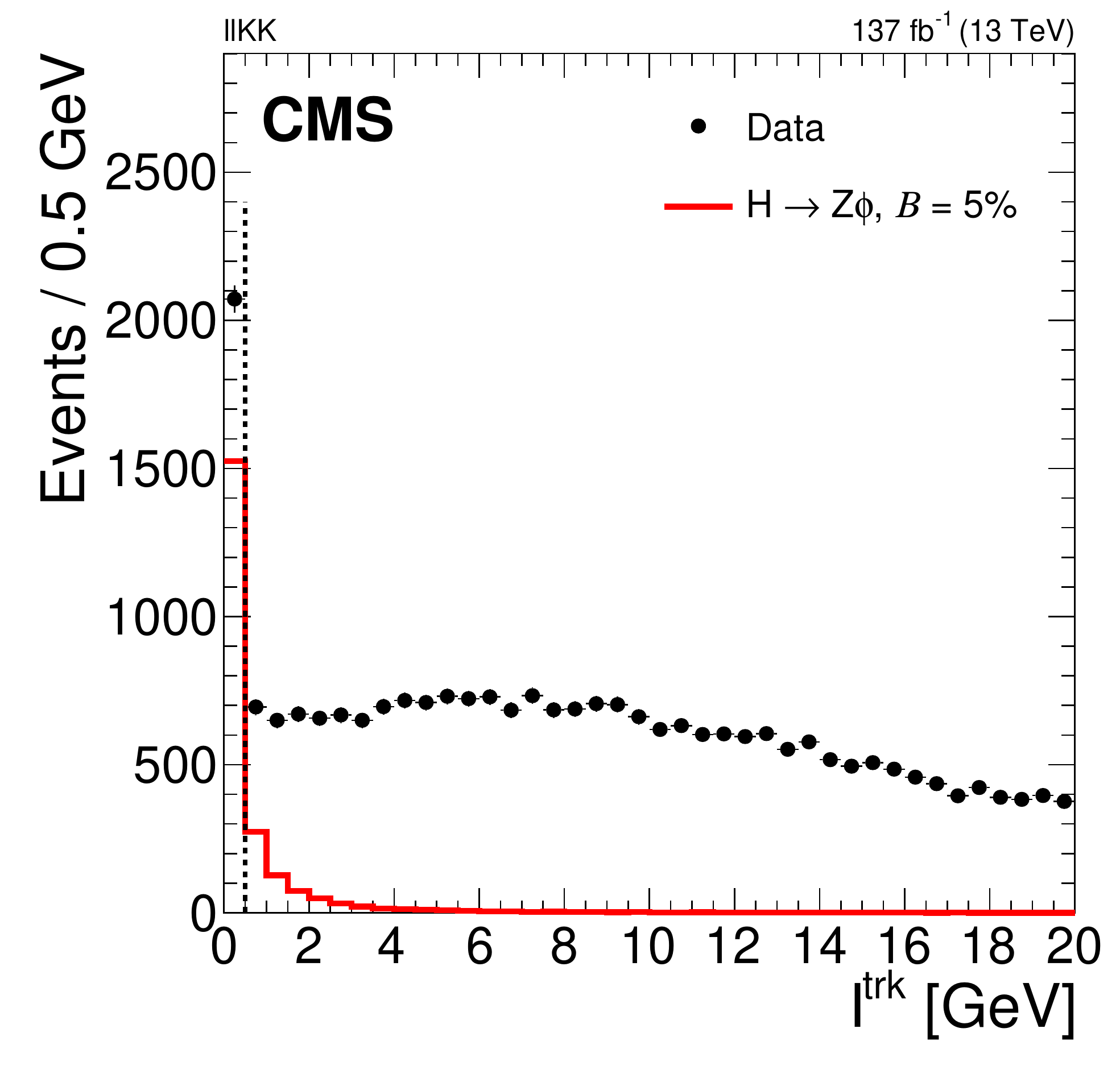}
\caption{The ditrack isolation sum in the $\llrho$ (left) and $\llphi$ (right) channels, combining the $\mumu$ and $\ee$ channels for all the data-taking years. The distribution in data, as well as in the simulated $\Hzrho$ and $\Hzphi$ signals is shown. A branching fraction of 10 (5)\% for the $\Hzrho$ ($\Hzphi$) signal is assumed. The isolation sum is shown after applying all selection criteria apart from the ditrack isolation requirement. The ditrack invariant mass requirement is also applied. Only events in which the dilepton plus ditrack invariant mass is in the range 120--130\GeV are considered. The dashed line indicates the boundary of
the region used in the analysis, for which the isolation sum is required to be smaller than 0.5\GeV.}
\label{fig:isolation_sum}
\end{figure}

Furthermore, we require the ditrack system to be isolated. An isolation sum $\Itrk$ is calculated as 
\begin{linenomath}
\begin{equation}
\Itrk=\sum \pt^{\text{trk}},
\end{equation}
\end{linenomath}
in a cone of radius $\Delta R = 0.3$ around the direction of the ditrack system. Only tracks with $\pt>0.5\GeV$ that are associated with the PV are considered, and the tracks forming the \PGr or \PGf candidate are excluded from the sum.
Events are selected if \mbox{$\Itrk < 0.5\GeV$}, thus with no track around the direction of the ditrack system. Figure \ref{fig:isolation_sum} shows the distributions of the isolation sum for the data and for the simulated signal, after applying all selection criteria except for the ditrack isolation requirement. The ditrack invariant mass requirement discussed below is also applied. This figure illustrates the reduction in background events due to the isolation requirements. Only events in which the dilepton and ditrack four-body mass is in the range 120--130\GeV are shown. This range is expected to contain 95\% of the simulated signal.

\begin{figure}[h!]
\centering
\includegraphics[width=0.48\textwidth]{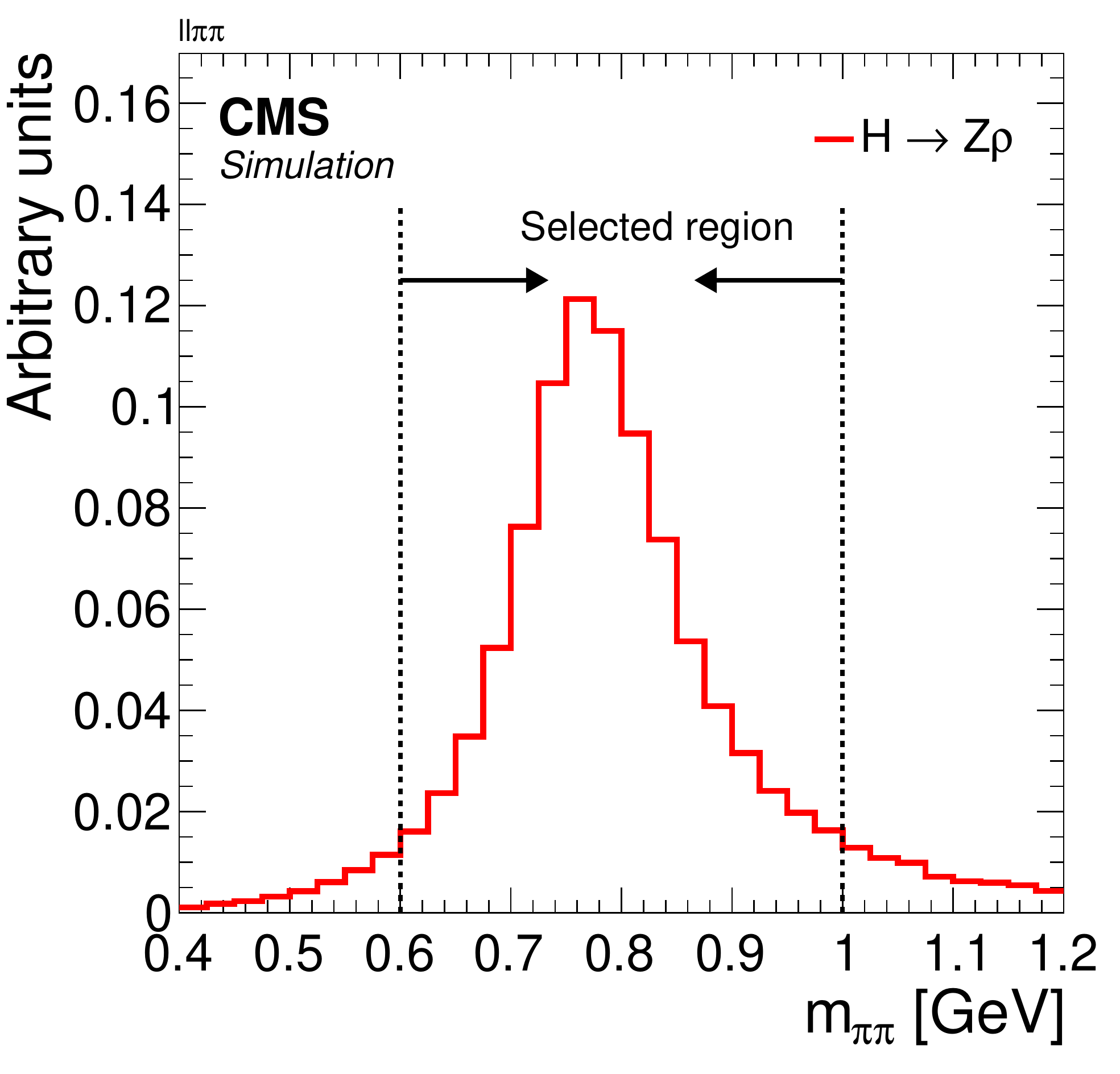}
\includegraphics[width=0.48\textwidth]{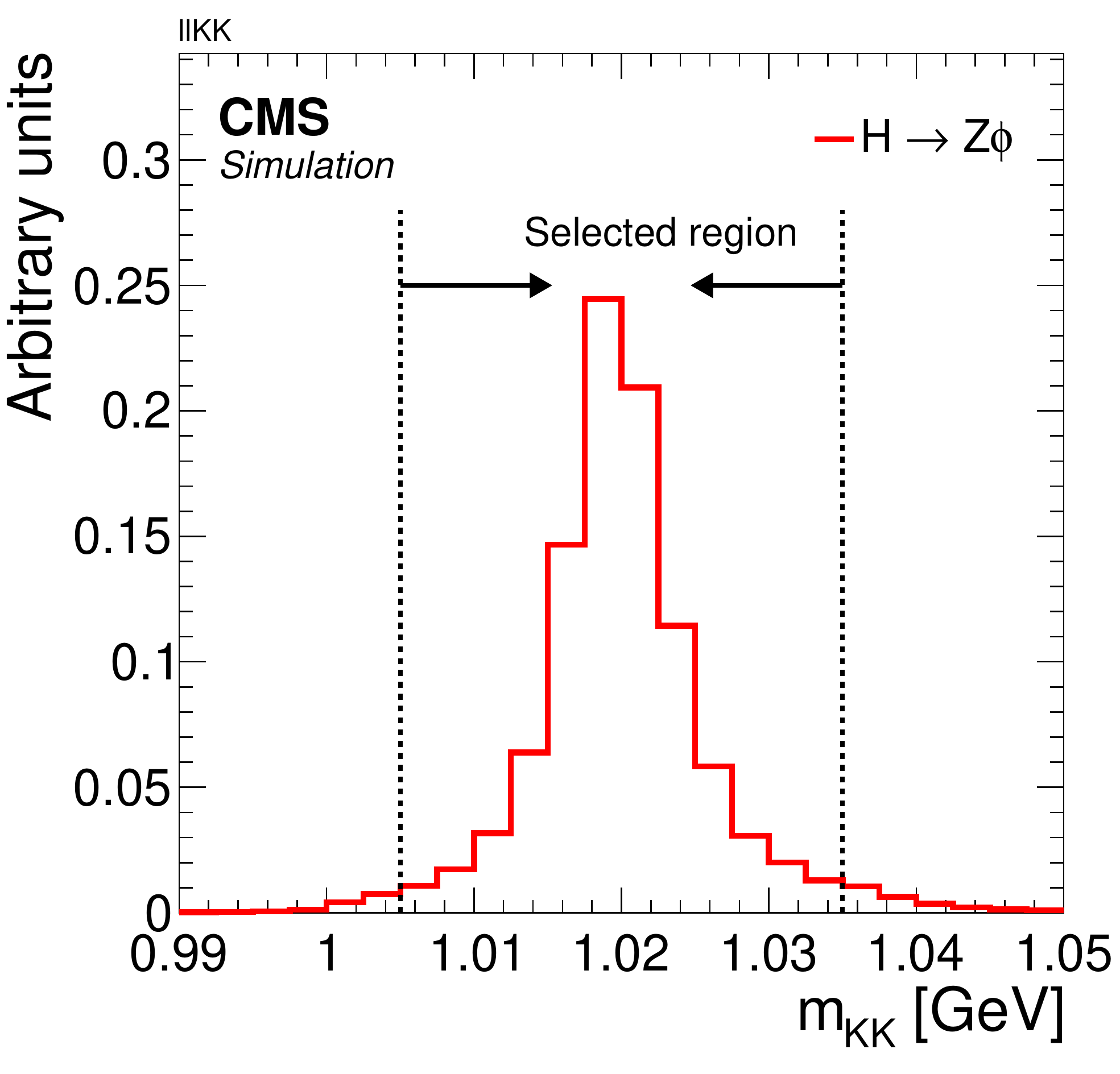}
\caption{Distribution of the ditrack invariant mass in simulated $\Hzrho$ events passing the $\llrho$ selection criteria (left) 
and in simulated $\Hzphi$ events passing the $\llphi$ selection criteria (right).
These masses are calculated assuming the charged particle mass equals the pion mass in the $\llrho$ selection and assuming the charged particle mass equals the kaon mass in the $\llphi$ selection.
The events pass all selection criteria described in the text, apart from the requirements on the ditrack invariant mass window. The dashed lines indicate the region selected in the analysis.}
\label{fig:ditrack_invariantmass}
\end{figure}

The invariant mass of the ditrack system is also used to reduce the contamination from background events.
Events with a \PGr candidate are selected if the invariant mass of the ditrack system is within $0.6<\mpipi<1\GeV$, calculated assuming the mass of each particle equals \mbox{$m_{\Pgppm}=139.6\MeV$~\cite{PhysRevD.98.030001}}. The full width at half the maximum 
of the $\mpipi$ distribution is approximately 120\MeV in the simulated signal. Figure~\ref{fig:ditrack_invariantmass} (left) shows this invariant mass distribution in simulated $\Hzrho$ events.
The \PGf meson has a smaller natural width than the \PGr meson, therefore it is possible to use a narrower mass window. The 
full width at half the maximum of the $\mkk$ distribution in simulated signal samples is
approximately 5\MeV. To select events with a \PGf candidate, the mass of each particle is taken as $m_{\PKpm}=493.7\MeV$~\cite{PhysRevD.98.030001} and
we require $1.005<\mkk<1.035\GeV$. Figure~\ref{fig:ditrack_invariantmass} (right) shows this invariant mass distribution in simulated $\Hzphi$ events.

After these requirements, including those on the ditrack invariant mass, the contribution from $\Hzphi$ events in the $\llrho$ channel is smaller than 1\% of the number of expected signal events in this channel when the SM branching fractions for $\Hzrho$ and $\Hzphi$ are considered. The same is true for contributions from $\Hzrho$ events in the
$\llphi$ channel. After all selection criteria are applied, there is no overlap in the events selected by the $\llrho$ and $\llphi$ channels.

The product of signal selection efficiency and acceptance ($\ea$) corresponds to the fraction of simulated signal events that pass the selection. To calculate these values we use the nominal simulated sample, in which the decays of the \PH and \PZ bosons are modelled isotropically. On average over the three data-taking years, $\ea$ in the $\mumurho$ ($\mumuphi$) channel is 15 (18)\%. For the $\eerho$ ($\eephi$) channel the average $\ea$ is 8 (10)\%. 

\section{Corrections applied to simulated samples}
\label{sec:sample-corrections}
A correction is applied to the simulated events such 
that the PU distribution in simulation reproduces this distribution in data~\cite{Sirunyan:2018nqx}. 
Corrections are also applied to the simulation to account for differences
in the efficiencies of the trigger selection; of the ditrack isolation requirement; 
and of the lepton reconstruction, 
identification, and isolation between simulated events and data.
These corrections, deviating from unity by a few percent, are measured
using the ``tag-and-probe'' method~\cite{Khachatryan:2010xn}. The ditrack
isolation efficiency correction is determined in $\PZ\to\Pgm\Pgm$ events using the tag-and-probe method. Here, 
the efficiency of the requirement on $\Itrk$ is measured for the probe muon. A systematic uncertainty,
described in Section~\ref{sec:uncertainties-inference}, is applied to account
for the difference between the phase space where the correction is measured
and where it is applied.
Energy scale corrections, which are smaller than 1\%, are applied to
the muons and electrons~\cite{Khachatryan:2015hwa,Sirunyan:2018}.

The event simulations model the decays of the \PH and \PZ bosons isotropically, and so do not take into account the impact of particle helicities.
However, as there are only a few possibilities for polarizations in the final decay products, we calculate the angular distributions
for extreme polarizations and reweight the signal events accordingly following the method described in Ref.~\cite{Broughton:2018wyd}.
The \PZ boson and the \PGr or \PGf meson can either both be transversely polarized or both be longitudinally polarized.
The two leptons always have opposite helicity in the rest frame of the \PZ boson.
For each possibility the distribution of the polar angle between one of the pions or kaons and the meson, and between 
one of the leptons and the \PZ boson, is evaluated analytically. The signal templates are weighted to both of these distributions simultaneously.
We ensure that the total normalization of the signal, before event selection, is preserved by the reweighting. However,
the reweighting modifies the distribution of the kinematic variables, in particular by changing the lepton \pt. Therefore
the reweighting reduces (increases) the fraction of signal events that pass the selection criteria in the transversely (longitudinally) polarized case, and so this affects the final results.
 The change of the signal yield in the two extreme polarizations, relative to the scenario with isotropic decays, is given in Table~\ref{tab:helicity_normalisation}.

\begin{table}[h!]
\centering
\topcaption{The effect on the signal yields of reweighting to the extreme polarization scenarios, described in more detail in the text, relative to the scenario with isotropic decays. The change in the fraction of signal events that pass the selection criteria affects the final results of the analysis.}
\label{tab:helicity_normalisation}

\begin{tabular}{p{0.28\textwidth}cccc}
\hline
Polarization state & \multicolumn{4}{c}{Effect on yield}\\
    &  $\mumurho$ & $\mumuphi$ & $\eerho$ & $\eephi$ \\
\hline
Longitudinally polarized & $+$16\% & $+$17\% & $+$23\% & $+$21\%\\
Transversely polarized & $-$8\% & $-$9\% &$-$11\% & $-$11\%\\
\hline
\end{tabular}
\end{table}

\section{Signal and background modelling}
\label{sec:background-modelling}
The dilepton and ditrack four-body mass distribution, corresponding to the reconstructed Higgs boson mass and denoted $\mllhad$, where $\mathrm{h}$ refers to \Pgp or \PK, is used in the statistical inference. The signal and background
are therefore modelled as a function of this observable in the range $118<\mllhad<168\GeV$. More than 95\% of the expected signal is contained in the range $120<\mllhad<130\GeV$; the large tail used at higher masses helps to improve the stability of the background parameterization. As a result of the kinematic selection on the leptons and the meson candidates, the four-body mass distribution for the background changes from rising to falling between $115<\mllhad<118\GeV$. For this reason the lower bound of the range is taken as $\mllhad=118\GeV$.
The full width at half the maximum of the $\mllhad$ distribution in samples of simulated signal events amounts to 2--3\GeV, depending on the channel considered.
The signal is described through a binned template, built from simulated events. Each bin has a width of 1\GeV in the four-body mass, which matches the binning used for the data.

The background to this search, consisting mainly of Drell--Yan events, is modelled using analytic functions. The values of the parameters of these analytic functions are obtained directly in the final signal extraction fit. Prior to the signal extraction fit we need to determine a set of functional forms that can parameterize the background in the different channels and data-taking years. Two sidebands, \mbox{$118<\mllhad<120\GeV$} and {$130<\mllhad<168\GeV$}, are used for this.
Because the sideband with $\mllhad<120\GeV$ is short, we verify that the chosen functional forms also describe the background in a control region where we require \mbox{$1<\Itrk<2\GeV$}. The fitted values of the 
function parameters in the control region are not required to be the same as those in the analysis phase space.
In the control region the full four-body mass range \mbox{$118<\mllhad<168\GeV$} is considered. 

Chebyshev polynomials are used to describe the backgrounds. The order used depends on the channel
and data-taking period, and ranges from 2 to 5. These orders are determined in the sidebands and the control regions described above using an F-test~\cite{Fisher:1922}. With this method we test whether a polynomial of order $n+1$ fits the data significantly better than a polynomial of order $n$. If this is not the case, the polynomial of order $n$ is selected. The results of the fit are shown in Section~\ref{sec:results}.

Alternative functions can be used to estimate the bias from the choice of a particular background parameterization. 
As alternatives we choose exponential functions, as well as a function of the form 
\begin{linenomath}
\begin{equation}
f(x) = ( 1-x )^{p_1} x^{-p_2-p_3\ln{x}},
\end{equation}
\end{linenomath}
where $x=m / \sqrt{s}$, and $p_i$ are parameters of the fit. Here, $m$ represents the four-body mass and $\sqrt{s}= 13\TeV$. Such a function has also been used in searches for dijet resonances~\cite{Sirunyan:2018xlo}. These alternative functional forms have a different shape from the nominal background model, but still fit the data in the sidebands well.

The possible bias from the choice of background parameterization is estimated by fitting the alternative function to the four-body mass sidebands. Pseudo-experiments are 
then drawn from this parameterization, and a signal expectation is added to each pseudo-data set. 
A maximum likelihood fit of the signal and background models to each pseudo-data set is performed using the nominal background model. 
This test is performed three times with branching fractions of 0, 2.5, and 5\% for $\Hzrho$ or $\Hzphi$. The test is also performed with both alternative functions described in the previous paragraph. 
The difference between the extracted and injected branching fraction is, within uncertainties, compatible between the tests with different injected branching fractions. This difference, for the alternative function for which it is largest, is taken as the uncertainty due to a possible bias in the choice of background parameterization. 
The bias is found to be small and is included in the analysis as a systematic uncertainty.

\section{Signal extraction and systematic uncertainties}
\label{sec:uncertainties-inference}
The results of this analysis are presented as upper limits on $\mathcal{B}(\Hzrho)$ and 
on $\mathcal{B}(\Hzphi)$. All limits quoted in what follows are set at the 95\% \CL. Limits are set using the modified frequentist
$\CLs$ criterion~\cite{CLS2,Read:2002hq}, in which the profile likelihood ratio modified for upper limits~\cite{CMS-NOTE-2011-005} is used as the test statistic. In the limit setting procedure
we make use of the asymptotic approximation~\cite{Cowan:2010js}.

Several systematic uncertainties are incorporated in the likelihood as nuisance
parameters. They are described in this section and summarized in Table~\ref{tab:systematics-comparison}.

Most of the systematic uncertainties affect only the normalization of the simulated signal templates:
\renewcommand{\theenumi}{\roman{enumi}.}
\begin{enumerate}
\item The uncertainties in the integrated luminosity measurements are, respectively, 2.5, 2.3, and 2.5\% for the 2016, 2017, and 2018 data-taking periods~\cite{CMS:2017sdi,CMS:2018elu,CMS:2019jhq}. 
\item Uncertainties in the muon identification, isolation, and trigger efficiency measurements arise from the method used to measure the efficiency, from the difference between the kinematic phase space in which the measurement is performed and where it is applied, and from the limited size of the simulated samples used for the measurement in simulation~\cite{Sirunyan:2018}. These uncertainties affect the normalisation of the simulated processes by $\approx$1\% for all the data-taking periods.
\item Uncertainties in the electron reconstruction, identification, and trigger efficiency measurements range from 2 to 3\%, depending on the data-taking period. These uncertainties mainly arise from the method used for the efficiency measurement~\cite{Khachatryan:2015hwa}.
\item The uncertainty in the tracking efficiency amounts to 4.6--4.8\% (corresponding to \mbox{2.3--2.4\%} per track), depending on the data-taking period. This uncertainty is determined by comparing ratios of \PDst meson decay chains in data and simulation. The dominant components of the uncertainty come from limited sample sizes and the uncertainties in the SM predictions of these ratios.
\item The uncertainty in the ditrack isolation efficiency measurement is 2\% for all three data-taking periods. This uncertainty mainly arises from the method used to measure the efficiency.
\item Theoretical uncertainties in the $\ggH$ production cross section amount to 3.9\%, with uncertainties in the VBF, $\WH$, and $\ZH$ production cross sections being, respectively, 0.4, 0.7, and 3.8\%~\cite{deFlorian:2016spz}.
\item Uncertainties from the choice of PDF and the value of the strong force coupling constant (\alpS) depend on the Higgs boson production mode and range from 1.6 to 3.2\%~\cite{deFlorian:2016spz}.
\end{enumerate}

Four systematic uncertainties affect both the shape and normalization of the simulated signal templates:
\begin{enumerate}
\item Uncertainties in the lepton energy scales are typically less than 0.3\% for both muons and electrons~\cite{Sirunyan:2018,Khachatryan:2015hwa}.
\item An additional uncertainty in the ditrack isolation efficiency measurement is applied. This uncertainty is taken as the difference between the ditrack isolation efficiency in the phase space where the correction is measured, and the efficiency as evaluated in the simulated signal. This uncertainty is in the range 1--6\%, depending on the data-taking period.
\item The uncertainty in the total inelastic cross section, used for correcting the PU profile in simulation to the profile in data, is 4.6\%~\cite{Sirunyan:2018nqx}. The overall effect on the normalisation of the simulated signal templates ranges from 0.5 to 1.5\%, depending on the data-taking period and the channel considered.
\item Uncertainties due to the limited number of simulated events are taken into account by allowing
each bin of the signal template to vary within its statistical uncertainty, independently
from the other bins.
\end{enumerate}

\begin{table}[h!]
\centering
\topcaption{Effect of systematic uncertainties on the simulated signal. The ranges reflect differences between channels and data-taking periods.}
\begin{tabular}{lll}
\hline
Uncertainty source & Type & Effect on simulated signal yield \\
\hline
Integrated luminosity & Normalization & 2.3--2.5\% \\
Muon efficiency & Normalization & 1\% \\
Muon energy scale & Shape & $<$0.3\%\\
Electron efficiency & Normalization & 2--3\% \\
Electron energy scale & Shape & $<$0.3\%\\
Tracking efficiency & Normalization & 4.6--4.8\%\\
Ditrack isolation efficiency & Normalization & 2\% \\
Ditrack isolation efficiency extrapolation & Shape & 1--6\%\\
Production cross sections & Normalization & 0.4--3.9\%\\
Choice of PDF and \alpS & Normalization & 1.6--3.2\%\\
Inelastic cross section & Shape & 0.5--1.5\%\\
Limited size of simulated samples & Shape & Bin-dependent\\
\hline
\end{tabular}
\label{tab:systematics-comparison}
\end{table}

The largest possible bias from the choice of the function modelling the background is included
in the likelihood as a modification of the number of expected events. The number
of expected events in a given bin $i$ is obtained as $(\mathcal{B}+\dbias)s_i+b_i$, 
where $s_i$ is the number of signal events and $b_i$ is the number of background events. The parameter $\mathcal{B}$ is
the branching fraction of the Higgs boson and the parameter on which we set limits. The parameter for the bias from the 
choice of background function is $\dbias$. It is subject to a Gaussian constraint
with a mean of 0 and a width equal to the largest possible bias due to the choice of background function, 
which ranges from 0.01 to 0.20\%. These values are obtained using the method described in Section~\ref{sec:background-modelling}.

Theoretical uncertainties in the production cross sections, and the uncertainties due to the choice of PDF and the value of \alpS are treated as correlated between the different data-taking periods. The uncertainty in the integrated luminosity measurement is treated as partially correlated between the different data-taking periods. The other experimental uncertainties are treated as uncorrelated between the different data-taking periods.

\section{Results}
\label{sec:results}
To present results in terms of $\mathcal{B}(\Hzrho)$ and  
$\mathcal{B}(\Hzphi)$, the signal templates are normalized by taking into
account the $\ggH$, VBF, $\WH$, and $\ZH$ production cross sections.
The $\ggH$ cross section is calculated at next-to-next-to-next-to-leading order in QCD and NLO in electroweak accuracy as 48.58\unit{pb}~\cite{deFlorian:2016spz}. The 
cross sections for the other production modes are calculated at next-to-next-to-leading order in QCD and NLO in electroweak accuracy, and amount, respectively, to
3.78, 1.37, and 0.88\unit{pb}~\cite{deFlorian:2016spz}. In addition, SM branching fractions of 3.37\% are assumed for each of the $\ztoll$ decays~\cite{PhysRevD.98.030001}.

In the limit setting procedure we do not take into account potential contributions of Higgs boson decays into a \PZ boson and other vector mesons. 

\begin{figure}[h!]
\centering
\includegraphics[width=0.48\textwidth]{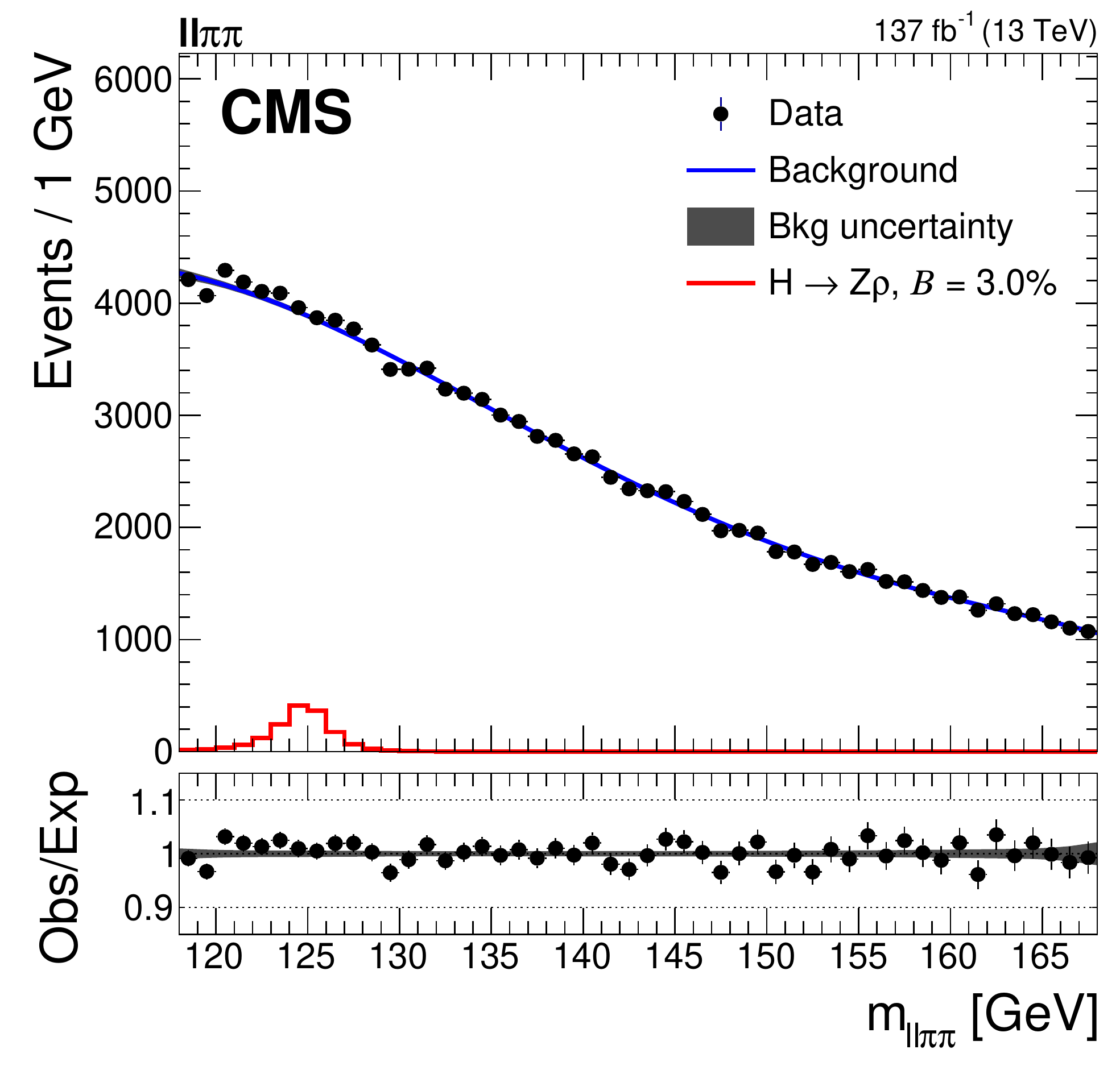}
\includegraphics[width=0.48\textwidth]{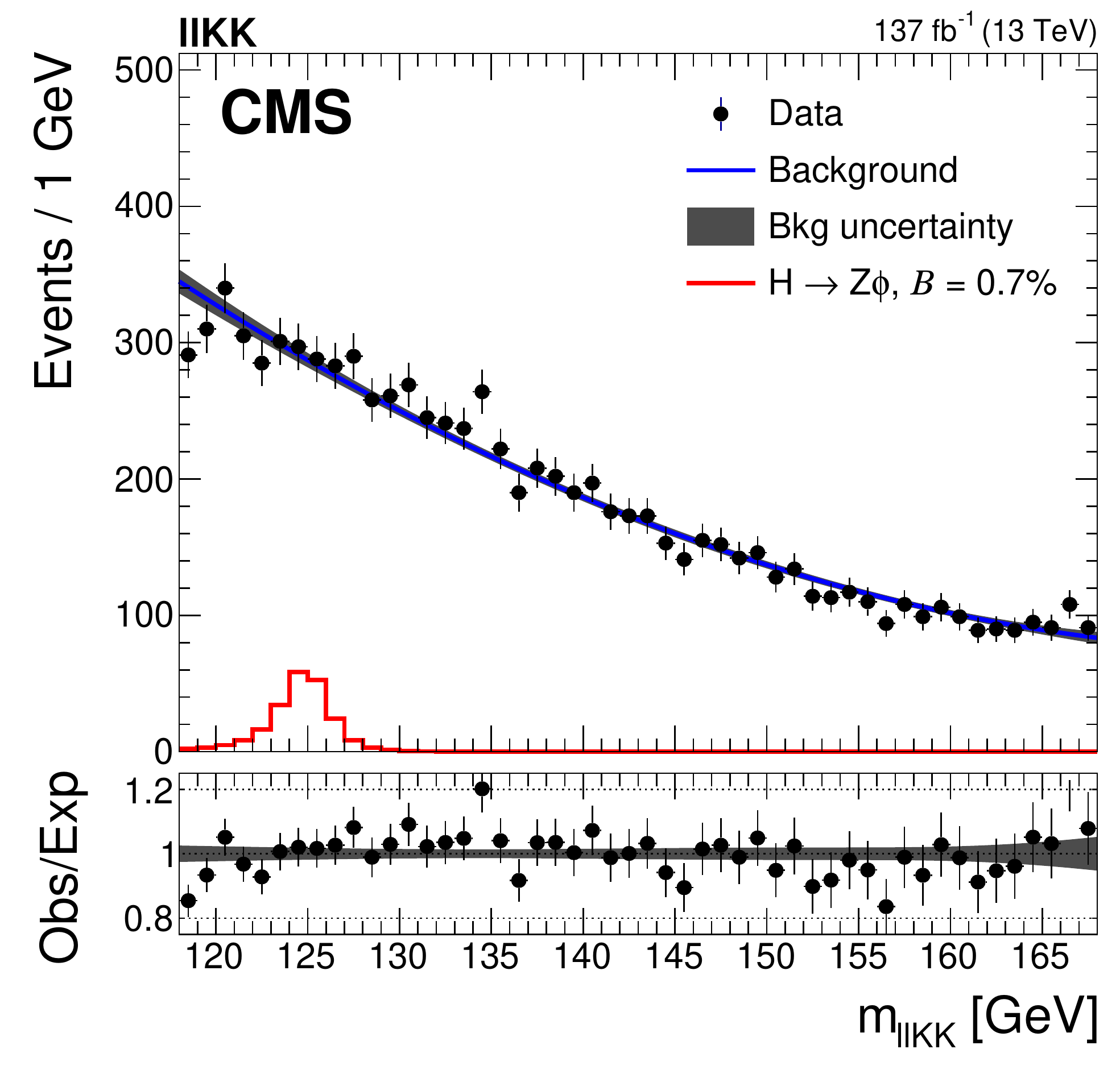}
\caption{Distributions of $m_{\llrho}$ (left) and $m_{\llphi}$ (right). For illustration the $\mumu$ and $\ee$ channels, as well as all three data-taking periods, are combined. Also shown are the $\Hzrho$ and $\Hzphi$ signals, in the isotropic-decay scenario and assuming branching fractions of 3.0 and 0.7\%, respectively. The ratio between the data and the background model is shown in the lower panels.}
\label{fig:sigplots}
\end{figure}

The four-body mass distributions in data and the background model are shown 
in Fig.~\ref{fig:sigplots}. The expected $\Hzrho$ ($\Hzphi$) signal, in the isotropic-decay scenario, at a branching fraction of 3.0 (0.7)\% is also shown. In this
figure the $\mumu$ and $\ee$ channels, as well as all three data-taking periods, are combined for illustration. In the statistical inference these
channels are considered separately in a simultaneous fit.
No significant excess above the background expectation is observed in either of the two searches. 

The observed upper limits on $\mathcal{B}(\Hzrho)$ and $\mathcal{B}(\Hzphi)$ are 1.04--1.31\% and 0.31--0.40\%, respectively, depending on the polarization scenario considered.
These values correspond to 740--940 times the SM expectation for the $\Hzrho$ decay and 730--950 times the SM expectation for the $\Hzphi$ decay.  
These limits can be compared with the expected upper limits, which are 0.63--0.80\% or 450--570 times the SM expectation for $\mathcal{B}(\Hzrho)$ and 0.27--0.36\% or 650--850 times the SM expectation for $\mathcal{B}(\Hzphi)$. These ranges reflect the considered polarization scenarios. The observed and expected upper limits are shown in Table~\ref{tab:limits-helicity-rho} for $\mathcal{B}(\Hzrho)$ and in Table~\ref{tab:limits-helicity-phi} for $\mathcal{B}(\Hzphi)$. While these limits are set on the total $\mathcal{B}(\Hzrho)$ and $\mathcal{B}(\Hzphi)$, 
the results mainly probe the indirect process via the $\Hzz$ decay as the direct decay process (Fig.~\ref{fig:diagrams}, right) is greatly suppressed in the SM.

\begin{table}[h!]
\centering
\topcaption{Observed and expected 95\% \CL upper limits on $\mathcal{B}(\Hzrho)$, for different polarizations.}
\cmsTable{
\begin{tabular}{lcccc}
\hline
  & Observed & Median expected & $\pm 68\%$ expected & $\pm 95\%$ expected\\
\hline
Isotropic decay & 1.21\% & 0.73\% & 0.52--1.04\% & 0.38--1.41\% \\ [\cmsTabSkip]
\PZ and \PGr longitudinally polarized & 1.04\% & 0.63\% & 0.44--0.89\% & 0.32--1.20\% \\
\PZ and \PGr transversely polarized & 1.31\% & 0.80\% & 0.57--1.14\% & 0.41--1.54\% \\
\hline
\end{tabular}
}
\label{tab:limits-helicity-rho}
\end{table}

\begin{table}[h!]
\centering
\topcaption{Observed and expected 95\% \CL upper limits on $\mathcal{B}(\Hzphi)$, for different polarizations.}
\cmsTable{
\begin{tabular}{lcccc}
\hline
  & Observed & Median expected & $\pm 68\%$ expected & $\pm 95\%$ expected\\
\hline
Isotropic decay & 0.36\% & 0.33\% & 0.23--0.46\% & 0.18--0.61\%\\ [\cmsTabSkip]
\PZ and \PGf longitudinally polarized & 0.31\% & 0.27\% & 0.20--0.39\% & 0.15--0.52\%\\
\PZ and \PGf transversely polarized & 0.40\% & 0.36\% & 0.26--0.50\% & 0.19--0.68\% \\
\hline
\end{tabular}
}
\label{tab:limits-helicity-phi}
\end{table}

\section{Summary}
\label{sec:summary}
A search has been presented for the rare decay of the Higgs boson (\PH) into a \PZ boson and a \PGr or a \PGf meson in the dilepton-$\pippim$ final states of the $\Hzrho$ decay, and in the dilepton-$\kpkm$ final states of the $\Hzphi$ decay. The search used a sample
of proton-proton collisions, collected by the CMS experiment at a centre-of-mass energy of 13\TeV from 2016 to 2018 and corresponding
to an integrated luminosity of 137\fbinv.
Upper limits on the branching fractions $\mathcal{B}(\Hzrho)$ and $\mathcal{B}(\Hzphi)$ have been set at the 95\% confidence level for various 
polarization scenarios. 
The upper limits on $\mathcal{B}(\Hzrho)$ are in the range 1.04--1.31\%, or 740--940 times the standard model expectation. The upper limits on $\mathcal{B}(\Hzphi)$ range from 0.31 to 0.40\%, or 730--950 times the standard model expectation. These results
constitute the first experimental limits on the two decay channels.

\begin{acknowledgments}
We congratulate our colleagues in the CERN accelerator departments for the excellent performance of the LHC and thank the technical and administrative staffs at CERN and at other CMS institutes for their contributions to the success of the CMS effort. In addition, we gratefully acknowledge the computing centres and personnel of the Worldwide LHC Computing Grid for delivering so effectively the computing infrastructure essential to our analyses. Finally, we acknowledge the enduring support for the construction and operation of the LHC and the CMS detector provided by the following funding agencies: BMBWF and FWF (Austria); FNRS and FWO (Belgium); CNPq, CAPES, FAPERJ, FAPERGS, and FAPESP (Brazil); MES (Bulgaria); CERN; CAS, MoST, and NSFC (China); COLCIENCIAS (Colombia); MSES and CSF (Croatia); RIF (Cyprus); SENESCYT (Ecuador); MoER, ERC IUT, PUT and ERDF (Estonia); Academy of Finland, MEC, and HIP (Finland); CEA and CNRS/IN2P3 (France); BMBF, DFG, and HGF (Germany); GSRT (Greece); NKFIA (Hungary); DAE and DST (India); IPM (Iran); SFI (Ireland); INFN (Italy); MSIP and NRF (Republic of Korea); MES (Latvia); LAS (Lithuania); MOE and UM (Malaysia); BUAP, CINVESTAV, CONACYT, LNS, SEP, and UASLP-FAI (Mexico); MOS (Montenegro); MBIE (New Zealand); PAEC (Pakistan); MSHE and NSC (Poland); FCT (Portugal); JINR (Dubna); MON, RosAtom, RAS, RFBR, and NRC KI (Russia); MESTD (Serbia); SEIDI, CPAN, PCTI, and FEDER (Spain); MOSTR (Sri Lanka); Swiss Funding Agencies (Switzerland); MST (Taipei); ThEPCenter, IPST, STAR, and NSTDA (Thailand); TUBITAK and TAEK (Turkey); NASU (Ukraine); STFC (United Kingdom); DOE and NSF (USA).

\hyphenation{Rachada-pisek} Individuals have received support from the Marie-Curie programme and the European Research Council and Horizon 2020 Grant, contract Nos.\ 675440, 752730, and 765710 (European Union); the Leventis Foundation; the A.P.\ Sloan Foundation; the Alexander von Humboldt Foundation; the Belgian Federal Science Policy Office; the Fonds pour la Formation \`a la Recherche dans l'Industrie et dans l'Agriculture (FRIA-Belgium); the Agentschap voor Innovatie door Wetenschap en Technologie (IWT-Belgium); the F.R.S.-FNRS and FWO (Belgium) under the ``Excellence of Science -- EOS" -- be.h project n.\ 30820817; the Beijing Municipal Science \& Technology Commission, No. Z191100007219010; the Ministry of Education, Youth and Sports (MEYS) of the Czech Republic; the Deutsche Forschungsgemeinschaft (DFG) under Germany's Excellence Strategy -- EXC 2121 ``Quantum Universe" -- 390833306; the Lend\"ulet (``Momentum") Programme and the J\'anos Bolyai Research Scholarship of the Hungarian Academy of Sciences, the New National Excellence Program \'UNKP, the NKFIA research grants 123842, 123959, 124845, 124850, 125105, 128713, 128786, and 129058 (Hungary); the Council of Science and Industrial Research, India; the HOMING PLUS programme of the Foundation for Polish Science, cofinanced from European Union, Regional Development Fund, the Mobility Plus programme of the Ministry of Science and Higher Education, the National Science Center (Poland), contracts Harmonia 2014/14/M/ST2/00428, Opus 2014/13/B/ST2/02543, 2014/15/B/ST2/03998, and 2015/19/B/ST2/02861, Sonata-bis 2012/07/E/ST2/01406; the National Priorities Research Program by Qatar National Research Fund; the Ministry of Science and Higher Education, project no. 02.a03.21.0005 (Russia); the Programa Estatal de Fomento de la Investigaci{\'o}n Cient{\'i}fica y T{\'e}cnica de Excelencia Mar\'{\i}a de Maeztu, grant MDM-2015-0509 and the Programa Severo Ochoa del Principado de Asturias; the Thalis and Aristeia programmes cofinanced by EU-ESF and the Greek NSRF; the Rachadapisek Sompot Fund for Postdoctoral Fellowship, Chulalongkorn University and the Chulalongkorn Academic into Its 2nd Century Project Advancement Project (Thailand); the Kavli Foundation; the Nvidia Corporation; the SuperMicro Corporation; the Welch Foundation, contract C-1845; and the Weston Havens Foundation (USA).\end{acknowledgments}

\bibliography{auto_generated}  
\cleardoublepage \appendix\section{The CMS Collaboration \label{app:collab}}\begin{sloppypar}\hyphenpenalty=5000\widowpenalty=500\clubpenalty=5000\input{HIG-19-012-authorlist.tex}\end{sloppypar}
%%% END EDITABLE REGION %%%
\end{document}

%% file: HIG-19-012-authorlist.tex
\vskip\cmsinstskip
\textbf{Yerevan Physics Institute, Yerevan, Armenia}\\*[0pt]
A.M.~Sirunyan$^{\textrm{\dag}}$, A.~Tumasyan
\vskip\cmsinstskip
\textbf{Institut f\"{u}r Hochenergiephysik, Wien, Austria}\\*[0pt]
W.~Adam, F.~Ambrogi, T.~Bergauer, M.~Dragicevic, J.~Er\"{o}, A.~Escalante~Del~Valle, R.~Fr\"{u}hwirth\cmsAuthorMark{1}, M.~Jeitler\cmsAuthorMark{1}, N.~Krammer, L.~Lechner, D.~Liko, T.~Madlener, I.~Mikulec, F.M.~Pitters, N.~Rad, J.~Schieck\cmsAuthorMark{1}, R.~Sch\"{o}fbeck, M.~Spanring, S.~Templ, W.~Waltenberger, C.-E.~Wulz\cmsAuthorMark{1}, M.~Zarucki
% ...
\vskip\cmsinstskip
\dag: Deceased\\
1:  Also at Vienna University of Technology, Vienna, Austria\\
% ...


%% file: HIG-19-012_temp.bbl
\providecommand{\href}[2]{#2}\begingroup\raggedright\begin{thebibliography}{10}%
\makeatletter
\providecommand{\hrefCMSnoop }[0]{\@secondoftwo}%
\makeatother
\providecommand{\doi}{\texttt{doi:}\begingroup \urlstyle{tt}\Url}

\bibitem{Aad:2012tfa}
\hrefCMSnoop {}{{ATLAS Collaboration}, ``Observation of a new particle in the
  search for the standard model {Higgs} boson with the {ATLAS} detector at the
  {LHC}'',} \textit{ Phys. Lett. B} \textbf{ 716} (2012) 1,
  \href{http://dx.doi.org/10.1016/j.physletb.2012.08.020}{\doi{10.1016/j.physletb.2012.08.020}},
\href{http://www.arXiv.org/abs/1207.7214}{\texttt{arXiv:1207.7214}}.
%%CITATION = ARXIV:1207.7214;%%.

\bibitem{Chatrchyan:2012xdj}
\hrefCMSnoop {}{{CMS Collaboration}, ``Observation of a new boson at a mass of
  125 {GeV} with the {CMS} experiment at the {LHC}'',} \textit{ Phys. Lett. B}
  \textbf{ 716} (2012) 30,
  \href{http://dx.doi.org/10.1016/j.physletb.2012.08.021}{\doi{10.1016/j.physletb.2012.08.021}},
\href{http://www.arXiv.org/abs/1207.7235}{\texttt{arXiv:1207.7235}}.
%%CITATION = ARXIV:1207.7235;%%.

\bibitem{Chatrchyan:2013lba}
\hrefCMSnoop {}{{CMS Collaboration}, ``Observation of a new boson with mass
  near 125 {GeV} in pp collisions at $\sqrt{s}$ = 7 and 8 {TeV}'',} \textit{
  JHEP} \textbf{ 06} (2013) 081,
  \href{http://dx.doi.org/10.1007/JHEP06(2013)081}{\doi{10.1007/JHEP06(2013)081}},
  \href{http://www.arXiv.org/abs/1303.4571}{\texttt{arXiv:1303.4571}}.

\bibitem{Englert:1964et}
\hrefCMSnoop {}{F.~Englert and R.~Brout, ``Broken symmetry and the mass of
  gauge vector mesons'',} \textit{ Phys. Rev. Lett.} \textbf{ 13} (1964) 321,
\href{http://dx.doi.org/10.1103/PhysRevLett.13.321}{\doi{10.1103/PhysRevLett.13.321}}.
%%CITATION = PRLTA,13,321;%%.

\bibitem{Higgs:1964ia}
\hrefCMSnoop {}{P.~W. Higgs, ``Broken symmetries, massless particles and gauge
  fields'',} \textit{ Phys. Lett.} \textbf{ 12} (1964) 132,
\href{http://dx.doi.org/10.1016/0031-9163(64)91136-9}{\doi{10.1016/0031-9163(64)91136-9}}.
%%CITATION = PHLTA,12,132;%%.

\bibitem{Higgs:1964pj}
\hrefCMSnoop {}{P.~W. Higgs, ``Broken symmetries and the masses of gauge
  bosons'',} \textit{ Phys. Rev. Lett.} \textbf{ 13} (1964) 508,
\href{http://dx.doi.org/10.1103/PhysRevLett.13.508}{\doi{10.1103/PhysRevLett.13.508}}.
%%CITATION = PRLTA,13,508;%%.

\bibitem{Guralnik:1964eu}
\hrefCMSnoop {}{G.~S. Guralnik, C.~R. Hagen, and T.~W.~B. Kibble, ``Global
  conservation laws and massless particles'',} \textit{ Phys. Rev. Lett.}
  \textbf{ 13} (1964) 585,
\href{http://dx.doi.org/10.1103/PhysRevLett.13.585}{\doi{10.1103/PhysRevLett.13.585}}.
%%CITATION = PRLTA,13,585;%%.

\bibitem{Higgs:1966ev}
\hrefCMSnoop {}{P.~W. Higgs, ``Spontaneous symmetry breakdown without massless
  bosons'',} \textit{ Phys. Rev.} \textbf{ 145} (1966) 1156,
\href{http://dx.doi.org/10.1103/PhysRev.145.1156}{\doi{10.1103/PhysRev.145.1156}}.
%%CITATION = PHRVA,145,1156;%%.

\bibitem{Kibble:1967sv}
\hrefCMSnoop {}{T.~W.~B. Kibble, ``Symmetry breaking in non-abelian gauge
  theories'',} \textit{ Phys. Rev.} \textbf{ 155} (1967) 1554,
\href{http://dx.doi.org/10.1103/PhysRev.155.1554}{\doi{10.1103/PhysRev.155.1554}}.
%%CITATION = PHRVA,155,1554;%%.

\bibitem{Aad:2014eha}
\hrefCMSnoop {}{{ATLAS Collaboration}, ``Measurement of {Higgs} boson
  production in the diphoton decay channel in pp collisions at center-of-mass
  energies of 7 and 8 {TeV} with the {ATLAS} detector'',} \textit{ Phys. Rev.
  D} \textbf{ 90} (2014) 112015,
  \href{http://dx.doi.org/10.1103/PhysRevD.90.112015}{\doi{10.1103/PhysRevD.90.112015}},
\href{http://www.arXiv.org/abs/1408.7084}{\texttt{arXiv:1408.7084}}.
%%CITATION = ARXIV:1408.7084;%%.

\bibitem{Khachatryan:2014ira}
\hrefCMSnoop {}{{CMS Collaboration}, ``Observation of the diphoton decay of the
  {Higgs} boson and measurement of its properties'',} \textit{ Eur. Phys. J. C}
  \textbf{ 74} (2014) 3076,
  \href{http://dx.doi.org/10.1140/epjc/s10052-014-3076-z}{\doi{10.1140/epjc/s10052-014-3076-z}},
\href{http://www.arXiv.org/abs/1407.0558}{\texttt{arXiv:1407.0558}}.
%%CITATION = ARXIV:1407.0558;%%.

\bibitem{Aad:2014eva}
\hrefCMSnoop {}{{ATLAS Collaboration}, ``Measurements of {Higgs} boson
  production and couplings in the four-lepton channel in pp collisions at
  center-of-mass energies of 7 and 8 {TeV} with the {ATLAS} detector'',}
  \textit{ Phys. Rev. D} \textbf{ 91} (2015) 012006,
  \href{http://dx.doi.org/10.1103/PhysRevD.91.012006}{\doi{10.1103/PhysRevD.91.012006}},
\href{http://www.arXiv.org/abs/1408.5191}{\texttt{arXiv:1408.5191}}.
%%CITATION = ARXIV:1408.5191;%%.

\bibitem{Chatrchyan:2013mxa}
\hrefCMSnoop {}{{CMS Collaboration}, ``Measurement of the properties of a
  {Higgs} boson in the four-lepton final state'',} \textit{ Phys. Rev. D}
  \textbf{ 89} (2014) 092007,
  \href{http://dx.doi.org/10.1103/PhysRevD.89.092007}{\doi{10.1103/PhysRevD.89.092007}},
\href{http://www.arXiv.org/abs/1312.5353}{\texttt{arXiv:1312.5353}}.
%%CITATION = ARXIV:1312.5353;%%.

\bibitem{ATLAS:2014aga}
\hrefCMSnoop {}{{ATLAS Collaboration}, ``Observation and measurement of {Higgs}
  boson decays to {WW$^*$} with the {ATLAS} detector'',} \textit{ Phys. Rev. D}
  \textbf{ 92} (2015) 012006,
  \href{http://dx.doi.org/10.1103/PhysRevD.92.012006}{\doi{10.1103/PhysRevD.92.012006}},
\href{http://www.arXiv.org/abs/1412.2641}{\texttt{arXiv:1412.2641}}.
%%CITATION = ARXIV:1412.2641;%%.

\bibitem{Aad:2015ona}
\hrefCMSnoop {}{{ATLAS Collaboration}, ``Study of {(W/Z)H} production and
  {Higgs} boson couplings using {$H \rightarrow WW^{\ast}$} decays with the
  {ATLAS} detector'',} \textit{ JHEP} \textbf{ 08} (2015) 137,
  \href{http://dx.doi.org/10.1007/JHEP08(2015)137}{\doi{10.1007/JHEP08(2015)137}},
\href{http://www.arXiv.org/abs/1506.06641}{\texttt{arXiv:1506.06641}}.
%%CITATION = ARXIV:1506.06641;%%.

\bibitem{Chatrchyan:2013iaa}
\hrefCMSnoop {}{{CMS Collaboration}, ``Measurement of {Higgs} boson production
  and properties in the {WW} decay channel with leptonic final states'',}
  \textit{ JHEP} \textbf{ 01} (2014) 096,
  \href{http://dx.doi.org/10.1007/JHEP01(2014)096}{\doi{10.1007/JHEP01(2014)096}},
\href{http://www.arXiv.org/abs/1312.1129}{\texttt{arXiv:1312.1129}}.
%%CITATION = ARXIV:1312.1129;%%.

\bibitem{Khachatryan:2016vau}
\hrefCMSnoop {}{{ATLAS and CMS Collaborations}, ``Measurements of the {Higgs}
  boson production and decay rates and constraints on its couplings from a
  combined {ATLAS} and {CMS} analysis of the {LHC} pp collision data at $
  \sqrt{s}=7 $ and 8 {TeV}'',} \textit{ JHEP} \textbf{ 08} (2016) 045,
  \href{http://dx.doi.org/10.1007/JHEP08(2016)045}{\doi{10.1007/JHEP08(2016)045}},
  \href{http://www.arXiv.org/abs/1606.02266}{\texttt{arXiv:1606.02266}}.

\bibitem{Aaboud:2018pen}
\hrefCMSnoop {}{{ATLAS Collaboration}, ``Cross-section measurements of the
  {Higgs} boson decaying into a pair of $\tau$-leptons in proton-proton
  collisions at $\sqrt{s}=13$ {TeV} with the {ATLAS} detector'',} \textit{
  Phys. Rev. D} \textbf{ 99} (2019) 072001,
  \href{http://dx.doi.org/10.1103/PhysRevD.99.072001}{\doi{10.1103/PhysRevD.99.072001}},
  \href{http://www.arXiv.org/abs/1811.08856}{\texttt{arXiv:1811.08856}}.

\bibitem{Sirunyan:2017khh}
\hrefCMSnoop {}{{CMS Collaboration}, ``Observation of the {Higgs} boson decay
  to a pair of $\tau$ leptons with the {CMS} detector'',} \textit{ Phys. Lett.
  B} \textbf{ 779} (2018) 283,
  \href{http://dx.doi.org/10.1016/j.physletb.2018.02.004}{\doi{10.1016/j.physletb.2018.02.004}},
  \href{http://www.arXiv.org/abs/1708.00373}{\texttt{arXiv:1708.00373}}.

\bibitem{Aaboud:2018urx}
\hrefCMSnoop {}{{ATLAS Collaboration}, ``Observation of {Higgs} boson
  production in association with a top quark pair at the {LHC} with the {ATLAS}
  detector'',} \textit{ Phys. Lett. B} \textbf{ 784} (2018) 173,
  \href{http://dx.doi.org/10.1016/j.physletb.2018.07.035}{\doi{10.1016/j.physletb.2018.07.035}},
  \href{http://www.arXiv.org/abs/1806.00425}{\texttt{arXiv:1806.00425}}.

\bibitem{Sirunyan:2018hoz}
\hrefCMSnoop {}{{CMS Collaboration}, ``Observation of
  $\mathrm{t\overline{t}}${H} production'',} \textit{ Phys. Rev. Lett.}
  \textbf{ 120} (2018) 231801,
  \href{http://dx.doi.org/10.1103/PhysRevLett.120.231801}{\doi{10.1103/PhysRevLett.120.231801}},
  \href{http://www.arXiv.org/abs/1804.02610}{\texttt{arXiv:1804.02610}}.

\bibitem{Aaboud:2018zhk}
\hrefCMSnoop {}{{ATLAS Collaboration}, ``Observation of {$H \rightarrow
  b\bar{b}$} decays and {$VH$} production with the {ATLAS} detector'',}
  \textit{ Phys. Lett. B} \textbf{ 786} (2018) 59,
  \href{http://dx.doi.org/10.1016/j.physletb.2018.09.013}{\doi{10.1016/j.physletb.2018.09.013}},
  \href{http://www.arXiv.org/abs/1808.08238}{\texttt{arXiv:1808.08238}}.

\bibitem{Sirunyan:2018kst}
\hrefCMSnoop {}{{CMS Collaboration}, ``Observation of {Higgs} boson decay to
  bottom quarks'',} \textit{ Phys. Rev. Lett.} \textbf{ 121} (2018) 121801,
  \href{http://dx.doi.org/10.1103/PhysRevLett.121.121801}{\doi{10.1103/PhysRevLett.121.121801}},
  \href{http://www.arXiv.org/abs/1808.08242}{\texttt{arXiv:1808.08242}}.

\bibitem{Aad:2020xfq}
\hrefCMSnoop {}{{ATLAS Collaboration}, ``A search for the dimuon decay of the
  standard model {Higgs} boson with the {ATLAS} detector'',} (2020).
  \href{http://www.arXiv.org/abs/2007.07830}{\texttt{arXiv:2007.07830}}.
  Submitted to \textit{Phys. Lett. B}.

\bibitem{Sirunyan:2018hbu}
\hrefCMSnoop {}{{CMS Collaboration}, ``Search for the {Higgs} boson decaying to
  two muons in proton-proton collisions at $\sqrt{s} = 13$ {TeV}'',} \textit{
  Phys. Rev. Lett.} \textbf{ 122} (2019) 021801,
  \href{http://dx.doi.org/10.1103/PhysRevLett.122.021801}{\doi{10.1103/PhysRevLett.122.021801}},
\href{http://www.arXiv.org/abs/1807.06325}{\texttt{arXiv:1807.06325}}.
%%CITATION = ARXIV:1807.06325;%%.

\bibitem{Aaboud:2018fhh}
\hrefCMSnoop {}{{ATLAS Collaboration}, ``Search for the decay of the {Higgs}
  boson to charm quarks with the {ATLAS} experiment'',} \textit{ Phys. Rev.
  Lett.} \textbf{ 120} (2018) 211802,
  \href{http://dx.doi.org/10.1103/PhysRevLett.120.211802}{\doi{10.1103/PhysRevLett.120.211802}},
\href{http://www.arXiv.org/abs/1802.04329}{\texttt{arXiv:1802.04329}}.
%%CITATION = ARXIV:1802.04329;%%.

\bibitem{Sirunyan:2019qia}
\hrefCMSnoop {}{{CMS Collaboration}, ``A search for the standard model {Higgs}
  boson decaying to charm quarks'',} \textit{ JHEP} \textbf{ 03} (2020) 131,
  \href{http://dx.doi.org/10.1007/JHEP03(2020)131}{\doi{10.1007/JHEP03(2020)131}},
  \href{http://www.arXiv.org/abs/1912.01662}{\texttt{arXiv:1912.01662}}.

\bibitem{Bodwin:2013gca}
\hrefCMSnoop {}{G.~T. Bodwin, F.~Petriello, S.~Stoynev, and M.~Velasco,
  ``{Higgs} boson decays to quarkonia and the {$H\bar{c}c$} coupling'',}
  \textit{ Phys. Rev. D} \textbf{ 88} (2013) 053003,
  \href{http://dx.doi.org/10.1103/PhysRevD.88.053003}{\doi{10.1103/PhysRevD.88.053003}},
  \href{http://www.arXiv.org/abs/1306.5770}{\texttt{arXiv:1306.5770}}.

\bibitem{Kagan:2014ila}
A.~L. Kagan\hrefCMSnoop {}{ {et~al.}, ``Exclusive window onto {Higgs} {Yukawa}
  couplings'',} \textit{ Phys. Rev. Lett.} \textbf{ 114} (2015) 101802,
  \href{http://dx.doi.org/10.1103/PhysRevLett.114.101802}{\doi{10.1103/PhysRevLett.114.101802}},
  \href{http://www.arXiv.org/abs/1406.1722}{\texttt{arXiv:1406.1722}}.

\bibitem{Alte:2016yuw}
\hrefCMSnoop {}{S.~Alte, M.~K{\"o}nig, and M.~Neubert, ``Exclusive weak
  radiative {Higgs} decays in the standard model and beyond'',} \textit{ JHEP}
  \textbf{ 12} (2016) 037,
  \href{http://dx.doi.org/10.1007/JHEP12(2016)037}{\doi{10.1007/JHEP12(2016)037}},
\href{http://www.arXiv.org/abs/1609.06310}{\texttt{arXiv:1609.06310}}.
%%CITATION = ARXIV:1609.06310;%%.

\bibitem{Sirunyan:2018fmm}
\hrefCMSnoop {}{{CMS Collaboration}, ``Search for rare decays of {Z} and
  {Higgs} bosons to {J$/\psi$} and a photon in proton-proton collisions at
  $\sqrt{s} =13$ {TeV}'',} \textit{ Eur. Phys. J. C} \textbf{ 79} (2019) 94,
  \href{http://dx.doi.org/10.1140/epjc/s10052-019-6562-5}{\doi{10.1140/epjc/s10052-019-6562-5}},
\href{http://www.arXiv.org/abs/1810.10056}{\texttt{arXiv:1810.10056}}.
%%CITATION = ARXIV:1810.10056;%%.

\bibitem{Aaboud:2018txb}
\hrefCMSnoop {}{{ATLAS Collaboration}, ``Searches for exclusive {Higgs} and
  {$Z$} boson decays into {$J/\psi\gamma$, $\psi(2S)\gamma$, and
  $\Upsilon(nS)\gamma$} at $\sqrt{s}=13$ {TeV} with the {ATLAS} detector'',}
  \textit{ Phys. Lett. B} \textbf{ 786} (2018) 134,
  \href{http://dx.doi.org/10.1016/j.physletb.2018.09.024}{\doi{10.1016/j.physletb.2018.09.024}},
\href{http://www.arXiv.org/abs/1807.00802}{\texttt{arXiv:1807.00802}}.
%%CITATION = ARXIV:1807.00802;%%.

\bibitem{Aaboud:2017xnb}
\hrefCMSnoop {}{{ATLAS Collaboration}, ``Search for exclusive {Higgs} and {$Z$}
  boson decays to $\phi\gamma$ and $\rho\gamma$ with the {ATLAS} detector'',}
  \textit{ JHEP} \textbf{ 07} (2018) 127,
  \href{http://dx.doi.org/10.1007/JHEP07(2018)127}{\doi{10.1007/JHEP07(2018)127}},
\href{http://www.arXiv.org/abs/1712.02758}{\texttt{arXiv:1712.02758}}.
%%CITATION = ARXIV:1712.02758;%%.

\bibitem{deFlorian:2016spz}
\hrefCMSnoop {}{{LHC Higgs Cross Section Working Group}, ``Handbook of {LHC}
  {H}iggs cross sections: 4. {D}eciphering the nature of the {H}iggs sector'',}
  CERN Report CERN-2017-002-M, 2016.
\newblock
  \href{http://dx.doi.org/10.23731/CYRM-2017-002}{\doi{10.23731/CYRM-2017-002}},
  \href{http://www.arXiv.org/abs/1610.07922}{\texttt{arXiv:1610.07922}}.

\bibitem{Isidori:2013cla}
\hrefCMSnoop {}{G.~Isidori, A.~V. Manohar, and M.~Trott, ``Probing the nature
  of the {Higgs}-like boson via $h \to v \mathcal{F}$ decays'',} \textit{ Phys.
  Lett. B} \textbf{ 728} (2014) 131,
  \href{http://dx.doi.org/10.1016/j.physletb.2013.11.054}{\doi{10.1016/j.physletb.2013.11.054}},
\href{http://www.arXiv.org/abs/1305.0663}{\texttt{arXiv:1305.0663}}.
%%CITATION = ARXIV:1305.0663;%%.

\bibitem{Aad:2020hzm}
\hrefCMSnoop {}{{ATLAS Collaboration}, ``Search for {Higgs} boson decays into a
  {$Z$} boson and a light hadronically decaying resonance using 13 {TeV} $pp$
  collision data from the {ATLAS} detector'',} (2020).
  \href{http://www.arXiv.org/abs/2004.01678}{\texttt{arXiv:2004.01678}}.
  Submitted to \textit{Phys. Rev. Lett.}

\bibitem{Giudice:2008uua}
\hrefCMSnoop {}{G.~F. Giudice and O.~Lebedev, ``{Higgs}-dependent {Yukawa}
  couplings'',} \textit{ Phys. Lett. B} \textbf{ 665} (2008) 79,
  \href{http://dx.doi.org/10.1016/j.physletb.2008.05.062}{\doi{10.1016/j.physletb.2008.05.062}},
\href{http://www.arXiv.org/abs/0804.1753}{\texttt{arXiv:0804.1753}}.
%%CITATION = ARXIV:0804.1753;%%.

\bibitem{Bishara:2015cha}
\hrefCMSnoop {}{F.~Bishara, J.~Brod, P.~Uttayarat, and J.~Zupan, ``Nonstandard
  {Yukawa} couplings and {Higgs} portal dark matter'',} \textit{ JHEP} \textbf{
  01} (2016) 010,
  \href{http://dx.doi.org/10.1007/JHEP01(2016)010}{\doi{10.1007/JHEP01(2016)010}},
\href{http://www.arXiv.org/abs/1504.04022}{\texttt{arXiv:1504.04022}}.
%%CITATION = ARXIV:1504.04022;%%.

\bibitem{Egana-Ugrinovic:2019dqu}
\hrefCMSnoop {}{D.~Egana-Ugrinovic, S.~Homiller, and P.~R. Meade, ``{Higgs}
  bosons with large couplings to light quarks'',} \textit{ Phys. Rev. D}
  \textbf{ 100} (2019) 115041,
  \href{http://dx.doi.org/10.1103/PhysRevD.100.115041}{\doi{10.1103/PhysRevD.100.115041}},
  \href{http://www.arXiv.org/abs/1908.11376}{\texttt{arXiv:1908.11376}}.

\bibitem{Froggatt:1978nt}
\hrefCMSnoop {}{C.~D. Froggatt and H.~B. Nielsen, ``Hierarchy of quark masses,
  {Cabibbo} angles and {CP} violation'',} \textit{ Nucl. Phys. B} \textbf{ 147}
  (1979) 277,
\href{http://dx.doi.org/10.1016/0550-3213(79)90316-X}{\doi{10.1016/0550-3213(79)90316-X}}.
%%CITATION = NUPHA,B147,277;%%.

\bibitem{Randall:1999ee}
\hrefCMSnoop {}{L.~Randall and R.~Sundrum, ``A large mass hierarchy from a
  small extra dimension'',} \textit{ Phys. Rev. Lett.} \textbf{ 83} (1999)
  3370,
  \href{http://dx.doi.org/10.1103/PhysRevLett.83.3370}{\doi{10.1103/PhysRevLett.83.3370}},
\href{http://www.arXiv.org/abs/hep-ph/9905221}{\texttt{arXiv:hep-ph/9905221}}.
%%CITATION = HEP-PH/9905221;%%.

\bibitem{Huber:2000ie}
\hrefCMSnoop {}{S.~J. Huber and Q.~Shafi, ``Fermion masses, mixings and proton
  decay in a {Randall-Sundrum} model'',} \textit{ Phys. Lett. B} \textbf{ 498}
  (2001) 256,
  \href{http://dx.doi.org/10.1016/S0370-2693(00)01399-X}{\doi{10.1016/S0370-2693(00)01399-X}},
  \href{http://www.arXiv.org/abs/hep-ph/0010195}{\texttt{arXiv:hep-ph/0010195}}.

\bibitem{TRK-11-001}
\hrefCMSnoop {}{{CMS Collaboration}, ``Description and performance of track and
  primary-vertex reconstruction with the {CMS} tracker'',} \textit{ JINST}
  \textbf{ 9} (2014) P10009,
  \href{http://dx.doi.org/10.1088/1748-0221/9/10/P10009}{\doi{10.1088/1748-0221/9/10/P10009}},
\href{http://www.arXiv.org/abs/1405.6569}{\texttt{arXiv:1405.6569}}.
%%CITATION = ARXIV:1405.6569;%%.

\bibitem{Sirunyan:2018}
\hrefCMSnoop {}{{CMS Collaboration}, ``Performance of the {CMS} muon detector
  and muon reconstruction with proton-proton collisions at {$\sqrt{s} = 13$
  TeV}'',} \textit{ JINST} \textbf{ 13} (2018) P06015,
  \href{http://dx.doi.org/10.1088/1748-0221/13/06/P06015}{\doi{10.1088/1748-0221/13/06/P06015}},
\href{http://www.arXiv.org/abs/1804.04528}{\texttt{arXiv:1804.04528}}.
%%CITATION = ARXIV:1804.04528;%%.

\bibitem{Khachatryan:2015hwa}
\hrefCMSnoop {}{{CMS Collaboration}, ``Performance of electron reconstruction
  and selection with the {CMS} detector in proton-proton collisions at
  {$\sqrt{s} = 8$ TeV}'',} \textit{ JINST} \textbf{ 10} (2015) P06005,
  \href{http://dx.doi.org/10.1088/1748-0221/10/06/P06005}{\doi{10.1088/1748-0221/10/06/P06005}},
\href{http://www.arXiv.org/abs/1502.02701}{\texttt{arXiv:1502.02701}}.
%%CITATION = ARXIV:1502.02701;%%.

\bibitem{Khachatryan:2016bia}
\hrefCMSnoop {}{{CMS Collaboration}, ``The {CMS} trigger system'',} \textit{
  JINST} \textbf{ 12} (2017) P01020,
  \href{http://dx.doi.org/10.1088/1748-0221/12/01/P01020}{\doi{10.1088/1748-0221/12/01/P01020}},
\href{http://www.arXiv.org/abs/1609.02366}{\texttt{arXiv:1609.02366}}.
%%CITATION = ARXIV:1609.02366;%%.

\bibitem{Chatrchyan:2008zzk}
\hrefCMSnoop {}{{CMS Collaboration}, ``The {CMS} experiment at the {CERN}
  {LHC}'',} \textit{ JINST} \textbf{ 3} (2008) S08004,
  \href{http://dx.doi.org/10.1088/1748-0221/3/08/S08004}{\doi{10.1088/1748-0221/3/08/S08004}}.

\bibitem{Sirunyan:2017ulk}
\hrefCMSnoop {}{{CMS Collaboration}, ``Particle-flow reconstruction and global
  event description with the {CMS} detector'',} \textit{ JINST} \textbf{ 12}
  (2017) P10003,
  \href{http://dx.doi.org/10.1088/1748-0221/12/10/P10003}{\doi{10.1088/1748-0221/12/10/P10003}},
\href{http://www.arXiv.org/abs/1706.04965}{\texttt{arXiv:1706.04965}}.
%%CITATION = ARXIV:1706.04965;%%.

\bibitem{Cacciari:2008gp}
\hrefCMSnoop {}{M.~Cacciari, G.~P. Salam, and G.~Soyez, ``{The anti-\kt jet
  clustering algorithm}'',} \textit{ JHEP} \textbf{ 04} (2008) 063,
  \href{http://dx.doi.org/10.1088/1126-6708/2008/04/063}{\doi{10.1088/1126-6708/2008/04/063}},
  \href{http://www.arXiv.org/abs/0802.1189}{\texttt{arXiv:0802.1189}}.

\bibitem{Cacciari:2011ma}
\hrefCMSnoop {}{M.~Cacciari, G.~P. Salam, and G.~Soyez, ``{FastJet} user
  manual'',} \textit{ Eur. Phys. J. C} \textbf{ 72} (2012) 1896,
  \href{http://dx.doi.org/10.1140/epjc/s10052-012-1896-2}{\doi{10.1140/epjc/s10052-012-1896-2}},
\href{http://www.arXiv.org/abs/1111.6097}{\texttt{arXiv:1111.6097}}.
%%CITATION = ARXIV:1111.6097;%%.

\bibitem{Nason:2004rx}
\hrefCMSnoop {}{P.~Nason, ``A new method for combining {NLO QCD} with shower
  {Monte Carlo} algorithms'',} \textit{ JHEP} \textbf{ 11} (2004) 040,
  \href{http://dx.doi.org/10.1088/1126-6708/2004/11/040}{\doi{10.1088/1126-6708/2004/11/040}},
\href{http://www.arXiv.org/abs/hep-ph/0409146}{\texttt{arXiv:hep-ph/0409146}}.
%%CITATION = HEP-PH/0409146;%%.

\bibitem{Frixione:2007vw}
\hrefCMSnoop {}{S.~Frixione, P.~Nason, and C.~Oleari, ``Matching {NLO QCD}
  computations with parton shower simulations: the {POWHEG} method'',} \textit{
  JHEP} \textbf{ 11} (2007) 070,
  \href{http://dx.doi.org/10.1088/1126-6708/2007/11/070}{\doi{10.1088/1126-6708/2007/11/070}},
\href{http://www.arXiv.org/abs/0709.2092}{\texttt{arXiv:0709.2092}}.
%%CITATION = ARXIV:0709.2092;%%.

\bibitem{Alioli:2008tz}
\hrefCMSnoop {}{S.~Alioli, P.~Nason, C.~Oleari, and E.~Re, ``{NLO Higgs} boson
  production via gluon fusion matched with shower in {POWHEG}'',} \textit{
  JHEP} \textbf{ 04} (2009) 002,
  \href{http://dx.doi.org/10.1088/1126-6708/2009/04/002}{\doi{10.1088/1126-6708/2009/04/002}},
\href{http://www.arXiv.org/abs/0812.0578}{\texttt{arXiv:0812.0578}}.
%%CITATION = ARXIV:0812.0578;%%.

\bibitem{Alioli:2010xd}
\hrefCMSnoop {}{S.~Alioli, P.~Nason, C.~Oleari, and E.~Re, ``A general
  framework for implementing {NLO} calculations in shower {Monte Carlo}
  programs: the {POWHEG BOX}'',} \textit{ JHEP} \textbf{ 06} (2010) 043,
  \href{http://dx.doi.org/10.1007/JHEP06(2010)043}{\doi{10.1007/JHEP06(2010)043}},
\href{http://www.arXiv.org/abs/1002.2581}{\texttt{arXiv:1002.2581}}.
%%CITATION = ARXIV:1002.2581;%%.

\bibitem{Nason:2009ai}
\hrefCMSnoop {}{P.~Nason and C.~Oleari, ``{NLO Higgs} boson production via
  vector-boson fusion matched with shower in {POWHEG}'',} \textit{ JHEP}
  \textbf{ 02} (2010) 037,
  \href{http://dx.doi.org/10.1007/JHEP02(2010)037}{\doi{10.1007/JHEP02(2010)037}},
\href{http://www.arXiv.org/abs/0911.5299}{\texttt{arXiv:0911.5299}}.
%%CITATION = ARXIV:0911.5299;%%.

\bibitem{Luisoni:2013kna}
\hrefCMSnoop {}{G.~Luisoni, P.~Nason, C.~Oleari, and F.~Tramontano,
  ``{HW$^{\pm}$/HZ} + 0 and 1 jet at {NLO} with the {POWHEG BOX} interfaced to
  {GoSam} and their merging within {MiNLO}'',} \textit{ JHEP} \textbf{ 10}
  (2013) 083,
  \href{http://dx.doi.org/10.1007/JHEP10(2013)083}{\doi{10.1007/JHEP10(2013)083}},
\href{http://www.arXiv.org/abs/1306.2542}{\texttt{arXiv:1306.2542}}.
%%CITATION = ARXIV:1306.2542;%%.

\bibitem{Alwall:2014hca}
J.~Alwall\hrefCMSnoop {}{ {et~al.}, ``The automated computation of tree-level
  and next-to-leading order differential cross sections, and their matching to
  parton shower simulations'',} \textit{ JHEP} \textbf{ 07} (2014) 079,
  \href{http://dx.doi.org/10.1007/JHEP07(2014)079}{\doi{10.1007/JHEP07(2014)079}},
\href{http://www.arXiv.org/abs/1405.0301}{\texttt{arXiv:1405.0301}}.
%%CITATION = ARXIV:1405.0301;%%.

\bibitem{Sjostrand:2014zea}
T.~Sj{\"o}strand\hrefCMSnoop {}{ {et~al.}, ``An introduction to {PYTHIA
  8.2}'',} \textit{ Comput. Phys. Commun.} \textbf{ 191} (2015) 159,
  \href{http://dx.doi.org/10.1016/j.cpc.2015.01.024}{\doi{10.1016/j.cpc.2015.01.024}},
\href{http://www.arXiv.org/abs/1410.3012}{\texttt{arXiv:1410.3012}}.
%%CITATION = ARXIV:1410.3012;%%.

\bibitem{Ball:2014uwa}
\hrefCMSnoop {}{{NNPDF} Collaboration, ``Parton distributions for the {LHC} run
  {II}'',} \textit{ JHEP} \textbf{ 04} (2015) 040,
  \href{http://dx.doi.org/10.1007/JHEP04(2015)040}{\doi{10.1007/JHEP04(2015)040}},
\href{http://www.arXiv.org/abs/1410.8849}{\texttt{arXiv:1410.8849}}.
%%CITATION = ARXIV:1410.8849;%%.

\bibitem{Ball:2017nwa}
\hrefCMSnoop {}{{NNPDF} Collaboration, ``Parton distributions from
  high-precision collider data'',} \textit{ Eur. Phys. J. C} \textbf{ 77}
  (2017) 663,
  \href{http://dx.doi.org/10.1140/epjc/s10052-017-5199-5}{\doi{10.1140/epjc/s10052-017-5199-5}},
\href{http://www.arXiv.org/abs/1706.00428}{\texttt{arXiv:1706.00428}}.
%%CITATION = ARXIV:1706.00428;%%.

\bibitem{Khachatryan:2015pea}
\hrefCMSnoop {}{{CMS Collaboration}, ``Event generator tunes obtained from
  underlying event and multiparton scattering measurements'',} \textit{ Eur.
  Phys. J. C} \textbf{ 76} (2016) 155,
  \href{http://dx.doi.org/10.1140/epjc/s10052-016-3988-x}{\doi{10.1140/epjc/s10052-016-3988-x}},
\href{http://www.arXiv.org/abs/1512.00815}{\texttt{arXiv:1512.00815}}.
%%CITATION = ARXIV:1512.00815;%%.

\bibitem{Sirunyan:2019dfx}
\hrefCMSnoop {}{{CMS Collaboration}, ``Extraction and validation of a new set
  of {CMS PYTHIA8} tunes from underlying-event measurements'',} \textit{ Eur.
  Phys. J. C} \textbf{ 80} (2020) 4,
  \href{http://dx.doi.org/10.1140/epjc/s10052-019-7499-4}{\doi{10.1140/epjc/s10052-019-7499-4}},
\href{http://www.arXiv.org/abs/1903.12179}{\texttt{arXiv:1903.12179}}.
%%CITATION = ARXIV:1903.12179;%%.

\bibitem{Agostinelli:2002hh}
\hrefCMSnoop {}{{{\GEANTfour}} Collaboration, ``{\GEANTfour}---a simulation
  toolkit'',} \textit{ Nucl. Instrum. Meth. A} \textbf{ 506} (2003) 250,
\href{http://dx.doi.org/10.1016/S0168-9002(03)01368-8}{\doi{10.1016/S0168-9002(03)01368-8}}.
%%CITATION = NUIMA,A506,250;%%.

\bibitem{PhysRevD.98.030001}
\hrefCMSnoop {}{{Particle Data Group}, M.~Tanabashi {et~al.}, ``Review of
  particle physics'',} \textit{ Phys. Rev. D} \textbf{ 98} (2018) 030001,
  \href{http://dx.doi.org/10.1103/PhysRevD.98.030001}{\doi{10.1103/PhysRevD.98.030001}}.

\bibitem{Sirunyan:2018nqx}
\hrefCMSnoop {}{{CMS Collaboration}, ``Measurement of the inelastic
  proton-proton cross section at $\sqrt{s}=13$ {TeV}'',} \textit{ JHEP}
  \textbf{ 07} (2018) 161,
  \href{http://dx.doi.org/10.1007/JHEP07(2018)161}{\doi{10.1007/JHEP07(2018)161}},
  \href{http://www.arXiv.org/abs/1802.02613}{\texttt{arXiv:1802.02613}}.

\bibitem{Khachatryan:2010xn}
\hrefCMSnoop {}{{CMS Collaboration}, ``Measurements of inclusive {$\PW$} and
  {$\PZ$} cross sections in {$\Pp\Pp$} collisions at {$\sqrt{s}=7$ TeV}'',}
  \textit{ JHEP} \textbf{ 01} (2011) 080,
  \href{http://dx.doi.org/10.1007/JHEP01(2011)080}{\doi{10.1007/JHEP01(2011)080}},
\href{http://www.arXiv.org/abs/1012.2466}{\texttt{arXiv:1012.2466}}.
%%CITATION = ARXIV:1012.2466;%%.

\bibitem{Broughton:2018wyd}
\href
  {https://s3.cern.ch/inspire-prod-files-7/74772a163630218ad644acbae41e32a8}{J.~H.
  Broughton, ``Searches for Rare Exclusive {Higgs} Boson Decays to a Meson and
  an Associated Photon with the {ATLAS} Detector''}.
\newblock PhD thesis, Birmingham U., 2018.

\bibitem{Fisher:1922}
\hrefCMSnoop {}{R.~D. Fisher, ``On the interpretation of $\chi^2$ from
  contingency tables, and the calculation of {P}'',} \textit{ J. Royal Stat.
  Soc.} \textbf{ 85} (1922) 87,
  \href{http://dx.doi.org/10.2307/2340521}{\doi{10.2307/2340521}}.

\bibitem{Sirunyan:2018xlo}
\hrefCMSnoop {}{{CMS Collaboration}, ``Search for narrow and broad dijet
  resonances in proton-proton collisions at $ \sqrt{s}=13 $ {TeV} and
  constraints on dark matter mediators and other new particles'',} \textit{
  JHEP} \textbf{ 08} (2018) 130,
  \href{http://dx.doi.org/10.1007/JHEP08(2018)130}{\doi{10.1007/JHEP08(2018)130}},
\href{http://www.arXiv.org/abs/1806.00843}{\texttt{arXiv:1806.00843}}.
%%CITATION = ARXIV:1806.00843;%%.

\bibitem{CLS2}
\hrefCMSnoop {}{T.~Junk, ``Confidence level computation for combining searches
  with small statistics'',} \textit{ Nucl. Instrum. Meth. A} \textbf{ 434}
  (1999) 435,
  \href{http://dx.doi.org/10.1016/S0168-9002(99)00498-2}{\doi{10.1016/S0168-9002(99)00498-2}},
\href{http://www.arXiv.org/abs/hep-ex/9902006}{\texttt{arXiv:hep-ex/9902006}}.
%%CITATION = HEP-EX/9902006;%%.

\bibitem{Read:2002hq}
\hrefCMSnoop {}{A.~L. Read, ``Presentation of search results: The
  {CL$_{\text{s}}$} technique'',} \textit{ J. Phys. G} \textbf{ 28} (2002)
  2693,
\href{http://dx.doi.org/10.1088/0954-3899/28/10/313}{\doi{10.1088/0954-3899/28/10/313}}.
%%CITATION = JPAGA,G28,2693;%%.

\bibitem{CMS-NOTE-2011-005}
\href {https://cds.cern.ch/record/1379837}{{The ATLAS Collaboration, The CMS
  Collaboration, The LHC Higgs Combination Group}, ``Procedure for the {LHC}
  {Higgs} boson search combination in {Summer} 2011'',} Technical Report
  CMS-NOTE-2011-005, ATL-PHYS-PUB-2011-11, 2011.

\bibitem{Cowan:2010js}
\hrefCMSnoop {}{G.~Cowan, K.~Cranmer, E.~Gross, and O.~Vitells, ``Asymptotic
  formulae for likelihood-based tests of new physics'',} \textit{ Eur. Phys. J.
  C} \textbf{ 71} (2011) 1554,
  \href{http://dx.doi.org/10.1140/epjc/s10052-011-1554-0}{\doi{10.1140/epjc/s10052-011-1554-0}},
  \href{http://www.arXiv.org/abs/1007.1727}{\texttt{arXiv:1007.1727}}.
[Erratum: \DOI{10.1140/epjc/s10052-013-2501-z}].
%%CITATION = ARXIV:1007.1727;%%.

\bibitem{CMS:2017sdi}
\href {https://cds.cern.ch/record/2257069}{{CMS Collaboration}, ``{CMS}
  luminosity measurements for the 2016 data taking period'',} CMS Physics
  Analysis Summary CMS-PAS-LUM-17-001, 2017.

\bibitem{CMS:2018elu}
\href {https://cds.cern.ch/record/2621960}{{CMS Collaboration}, ``{CMS}
  luminosity measurement for the 2017 data-taking period at {$\sqrt{s} = 13$
  TeV}'',} CMS Physics Analysis Summary CMS-PAS-LUM-17-004, 2018.

\bibitem{CMS:2019jhq}
\href {https://cds.cern.ch/record/2676164}{{CMS Collaboration}, ``{CMS}
  luminosity measurement for the 2018 data-taking period at {$\sqrt{s} = 13$
  TeV}'',} CMS Physics Analysis Summary CMS-PAS-LUM-18-002, 2019.

\end{thebibliography}\endgroup
